\documentclass[twocolumn,secnumarabic,amsmath,amssymb, nobibnotes, aps, prb]{revtex4-1}
\usepackage{bmpsize}
\usepackage{graphicx}
\usepackage{dcolumn}
\usepackage{bm}
\usepackage{physics}
\usepackage{braket} 
\usepackage{hyperref}
\usepackage{alphabeta}
\usepackage{amsmath}
\usepackage{lipsum} 

\usepackage{booktabs}
\usepackage{alphabeta}
\usepackage{xcolor}
\def\SPSB#1#2{\rlap{\textsuperscript{#1}}{\textsubscript{#2}}}
\usepackage[english]{babel}
\usepackage{blindtext}
\usepackage{verbatim}

\begin{document}

\title{All-electron density functional calculations for electron and nuclear spin interactions in molecules and solids}

\author{Krishnendu Ghosh$^1$, He Ma$^2$$^,$$^3$, Vikram Gavini$^1$$^,$$^4$, and Giulia Galli$^2$$^,$$^3$$^,$$^5$}
\email{gagalli@uchicago.edu}
\affiliation{$^1$Department of Mechanical Engineering, University of Michigan, Ann Arbor MI 48109}
\affiliation{$^2$The Institute for Molecular Engineering, University of Chicago, Chicago IL 60637}
\affiliation{$^3$Department of Chemistry, University of Chicago, Chicago IL 60637}
\affiliation{$^4$Department of Materials Science and Engineering, University of Michigan, Ann Arbor MI 48109}
\affiliation{$^5$Materials Science Division, Argonne National Laboratory, Lemont IL 60439}

\date{\today}
\begin{abstract}
The interaction between electronic and nuclear spins in the presence of external magnetic fields can be described by a spin Hamiltonian, with parameters obtained from first principles, electronic structure calculations. We describe an approach to compute these parameters, applicable to both molecules and solids, which is based on Density Functional Theory (DFT) and real-space, all-electron calculations using finite elements (FE). We report results for hyperfine tensors, zero field splitting tensors (spin-spin component) and nuclear quadrupole tensors of a series of molecules and of the nitrogen-vacancy center in diamond. We compare our results with those of calculations using Gaussian orbitals and plane-wave basis sets, and we discuss their numerical accuracy. We show that calculations based on FE can be systematically converged with respect to the basis set, thus allowing one to establish reference values for the spin Hamiltonian parameters, at a given level of DFT. 
\end{abstract}
\maketitle

\section{Introduction} \label{Sec 1}
Electron spins in molecules, nanostructures and solids are important resources in many areas including spintronics~\cite{Zutic2004} and quantum information science~\cite{Weber2010}. For instance, high-spin magnetic molecules can be utilized as single-molecule magnets and are promising platforms for next-generation data storage devices~\cite{Guo2018}; in the solid state, spin-carrying deep centers in semiconductors can serve as quantum bits for quantum information processing~\cite{Koehl2015}. In order to understand  the physical properties of electron spins in molecules and solids, one needs to describe the interaction of electron and nuclear spins, in the presence of external electromagnetic fields. Such a description may be achieved by using spin Hamiltonians, with parameters derived from experiments or from calculations. For systems with a single effective electron spin, the leading terms in the spin Hamiltonian are~\cite{Schweiger2001, Harriman2013, Abragam2013}:
\begin{widetext}
\begin{equation} \label{Eq. 1}
H = \mu_B \bm{B} \cdot \bm{g} \cdot \bm{S} + \sum_N \gamma_N \bm{B} \cdot \bm{I}_N + \sum_N \bm{S} \cdot \bm{A}_N \cdot \bm{I}_N + \bm{S} \cdot \bm{D} \cdot \bm{S} + \sum_N \bm{I}_N \cdot \bm{P}_N \cdot \bm{I}_N
\end{equation}
\end{widetext}
where $\mu_B$ is Bohr magneton; $\bm{S}$ is the effective electron spin; $\bm{B}$ is the external magnetic field; $\bm{I}_N$ and $\gamma_N$ are the spin and gyromagnetic ratio of the $N^{\text{th}}$ nucleus; $\bm{g}$, $\bm{A}$, $\bm{D}$, and $\bm{P}$ are rank-2 tensors that characterize the strength of electron Zeeman interaction, hyperfine interaction, zero-field splitting and nuclear quadrupole interaction, respectively. Nuclear spin-spin interactions and the chemical shielding effect in nuclear Zeeman interactions are neglected in Eq. \ref{Eq. 1}.

The spin Hamiltonian parameters $\bm{g}$, $\bm{A}$, $\bm{D}$ and $\bm{P}$ may be obtained by electron paramagnetic resonance (EPR), nuclear quadrupole resonance (NQR) and related spectroscopic techniques~\cite{Weil2007}. Theoretically their values can be determined by first-principles electronic structure calculations, which also provide important information complementary to experiments. For example, in the case of spin defects in solids often times the atomistic structure and charge state of the defect are not straightforward to determine, experimentally. Comparing the computed spin Hamiltonian parameters for candidate structures and charge states with experimental results is a useful means to identify the properties of the defect. In addition, first-principles calculations can provide insights into the structure-property relations of molecules and spin defects, thus facilitating the rational design of molecules and materials with desirable spin properties. Finally, by simulating spin systems under external perturbations such as mechanical strain or applied electromagnetic fields, one can obtain valuable information and guidance for the experimental manipulation of electron spins \cite{Falk2014, Whiteley2018}.

Therefore, in order to devise predictive computational strategies, the development of robust methods for the calculation of spin Hamiltonian parameters is an important task. In spite of important progress in the fields of materials science~\cite{Vandewalle1993, Blugel1987, Pickard2001, Pickard2002, Bahramy2007, Rayson2008, Bodrog2013, Biktagirov2018} and quantum chemistry~\cite{Olsen2002, Latosinska2003, Sinnecker2006, Neese2005, Reviakine2006, Kossmann2007, Neese2007}, there is not yet a general and well established computational protocol that can reliably predict various spin Hamiltonian parameters with high accuracy for broad classes of systems. At present, the method most often adopted for spin Hamiltonian parameter calculations is Density Functional Theory (DFT). While calculations using \textit{ab initio} wavefunction-based methods have also appeared in the literature~\cite{Sayfutyarova2018, Sugisaki2009}, they have so far been limited to relatively small molecular systems due to their high computational cost. To solve the Kohn-Sham equations in DFT, single particle electronic wavefunctions are usually represented using basis sets, with Gaussian-type orbitals (GTO) and plane-waves (PW) being among the most popular choices for molecular and extended systems, respectively. In PW-based DFT calculations, pseudopotentials are employed and the electronic wavefunctions near the nuclei are not explicitly evaluated, and one generally needs to perform a so-called projected augmented wave (PAW) reconstruction~\cite{Blochl1994} to extract all-electron wavefunctions for the calculation of certain spin Hamiltonian parameters. Besides PW, there are studies exploring other basis sets including numerical atomic orbitals~\cite{Kadantsev2008}, linearized augmented plane-wave~\cite{Schwarz2003}, linear muffin-tin orbitals~\cite{Daalderop1996, Overhof2004}, and GTO~\cite{Dovesi2018} for the calculation of $A$-tensors and $V$-tensors (electric field gradient tensor, see Section II.3) for solids.

In this work we present calculations of spin Hamiltonian parameters carried out, for the first time, using a real-space finite-element (FE) formulation of DFT~\cite{Motamarri2013, dftfe}. The FE basis is a piece-wise continuous polynomial basis~\cite{Brenner2008}  that allows for systematic convergence of calculations with increasing polynomial order and decreasing element size. An important attribute of the FE basis is its spatial adaptivity that can provide increased resolution in specific regions of interest in real space, while using coarser descriptions elsewhere. In the present context, the FE basis can be chosen to have higher resolution in the core region to accurately describe the highly oscillatory nature of the single particle wavefunctions, and a coarser resolution far from the core where the orbitals are smoother. Further, FE-based calculations can be performed with either open or periodic boundary conditions, and therefore molecular and extended systems can be treated on an equal footing. There are several advantages in using FE-based DFT calculations for computing spin Hamiltonian parameters. The cusp of the wavefunctions near the nuclei can be more efficiently represented than with GTO basis sets, and this is an important requisite to compute quantities such as the Fermi contact component of the $A$-tensor. In addition, FE-based calculations can be systematically converged with respect to the basis set size in a more straightforward manner than GTO-based calculations, and they do not mandate the use of pseudopotentials and PAW reconstructions, as required when using PWs. 

The rest of the paper is organized as follows. In Section II we present the formalism for computing the spin Hamiltonian parameters in Eq. \ref{Eq. 1}, except those involving spin-orbit coupling contributions. Specifically, we consider the isotropic (Fermi contact) and the spin dipolar contribution to the hyperfine $A$-tensor, the spin-spin component of the zero-field splitting $D$-tensor, and the nuclear quadrupole $P$-tensor. The calculation of the $g$-tensor and the spin-orbit contributions to the $A$- and $D$-tensors will be subjects for future studies. In Section III we discuss our results for both molecules and solids, and Section IV concludes the paper.

\section{Real space computation of spin Hamiltonian parameters} \label{Sec 2}
The adaptivity of the FE basis to accurately and efficiently describe the all-electron Kohn-Sham orbitals is determined by the size of finite elements and the order of the polynomial basis functions. The FE mesh can be chosen adaptively (\textit{h}-refinement), so as to assign  smaller elements to regions requiring higher resolution (e.g.  around the nuclei) and coarser elements elsewhere. In addition, for a given  mesh, the order of the polynomial basis functions can be chosen (\textit{p}-refinement) to provide a high order function approximation. While both refinement approaches can be used to obtain systematic convergence, \textit{h}-refinement is suitable for realizing spatial adaptivity, and \textit{p}-refinement provides the flexibility to realize higher accuracy over the complete simulation domain.

\subsection{A-tensor}

The isotropic (Fermi contact) and the spin dipolar component of the $A$-tensor are given by:
\begin{equation} \label{Eq. 2} 
    A^{\text{iso}} = - \frac{1}{3S} \mu_0 \gamma_e \gamma_N \hbar^2 n_s(\bm{r}_N),
\end{equation}
\begin{widetext}
\begin{equation} \label{Eq. 3}
    A^{\text{dip}}_{ab} = \frac{1}{2S} \frac{\mu_0}{4\pi} \gamma_e \gamma_N \hbar^2 \int \frac{|\bm{r} - \bm{r}_N|^2 \delta_{ab} - 3 (\bm{r} - \bm{r}_N)_a (\bm{r} - \bm{r}_N)_b}{|\bm{r} - \bm{r}_N|^5} n_s(\bm{r}) d\bm{r},
\end{equation}
\end{widetext}
where $a, b = x, y, z$, S is the effective electron spin ($S=0$ for a singlet, $ \frac{1}{2}$ for a doublet, etc.); $n_s$ is the electron spin density; $\bm{r}_N$ is the position of the nucleus; $(\bm{r} - \bm{r}_N)_a$ is the $a$-direction component of $\bm{r} - \bm{r}_N$; $\gamma_e$ and $\gamma_N$ are gyromagnetic ratios for electron and nuclei, respectively. $\gamma_N$ for various nuclear isotopes can be obtained from standard databases, e.g. Ref. \onlinecite{WebElements}.

As can be seen from Eq. \ref{Eq. 2}, the isotropic (Fermi contact) component of the $A$-tensor exhibits a strong dependence on the electron spin density at the nuclei. An all-electron A-tensor calculation in real-space requires very refined finite elements near the nuclei to accurately compute the electron spin density. The spatial adaptivity of the finite element mesh (h refinement) is hence extremely useful here. On the other hand, the dipolar component of the A-tensor involves an integration with $\frac{1}{r^{5}}$ and $\frac{1}{r^{3}}$ kernels. This requires high accuracy in the electronic spin density within a certain region surrounding the nuclei, which can be systematically improved through the \textit{p}-refinement. 

\subsection{D-tensor}
The spin-spin component of the $D$-tensor evaluated using the Kohn-Sham wavefunctions, is given by~\cite{McWeeny1961, Rayson2008} 
\begin{widetext}
\begin{equation} \label{Eq. 4}
    D_{ab} = \frac{1}{2S(2S-1)} \frac{\mu_0}{4\pi} (\gamma_e \hbar)^2  \left[ \sum_{i < j}^{occ.} \chi_{ij} \int{\int{\Phi_{ij}^{*}(\textbf{r},\textbf{r}^{\ensuremath{'}}) \frac{ \tilde{r}^2\delta_{ab} - 3\tilde{r}_a \tilde{r}_b }{ r^5 }  \Phi_{ij}(\textbf{r},\textbf{r}^{\ensuremath{'}}) d\textbf{r} d\textbf{r}^{\ensuremath{'}}}} \right]\,,
\end{equation}
\end{widetext}
where the summation is over all pairs of occupied orbitals, and $\Phi_{ij}(\textbf{r},\textbf{r}^{\ensuremath{'}})$ are $2 \times 2$ determinants formed from orbitals $\phi_{i}$ and $\phi_{j}$, $\Phi_{ij}(\textbf{r},\textbf{r}^{\ensuremath{'}})=\frac{1}{\sqrt{2}}\Big[\phi_{i}(\textbf{r})\phi_{j}(\textbf{r}^{\ensuremath{'}}) - \phi_{i}(\textbf{r}^{\ensuremath{'}})\phi_{j}(\textbf{r})]$; $\chi_{ij} = \pm 1$ for parallel and antiparallel spins respectively; $\tilde{r}$ is a scalar representing $|\textbf{r}-\textbf{r}^{\ensuremath{'}}|$; $\tilde{r}_a$ represents the $a$-direction component of the vector $\textbf{r}-\textbf{r}^{\ensuremath{'}}$.
The operator $ \frac{ \tilde{r}^2\delta_{ab} - 3\tilde{r}_a \tilde{r}_b }{ r^5 } $ is the $ab$ element of the Hessian of the Green's function of $-\frac{1}{4\pi}\nabla^{2}$, i.e. $G(\textbf{r},\textbf{r}^{\ensuremath{'}})=\frac{1}{|\textbf{r}-\textbf{r}^{\ensuremath{'}}|}$. Since the operator, $\frac{\partial^{2}G(\textbf{r},\textbf{r}^{\ensuremath{'}})}{\partial r_a \partial r^{\ensuremath{'}}_b}$, is invariant under particle exchange, the real-space integrals in Eq. \ref{Eq. 4} can be split into direct ($M^{ij,D}_{ab}$) and exchange terms ($M^{ij,E}_{ab}$) given by 
\begin{equation} \label{Eq. 5}
    M^{ij,D}_{ab} = \int{\int{\phi_{i}(\textbf{r})\phi_{j}(\textbf{r}^{\ensuremath{'}})  \frac{\partial^{2}G(\textbf{r},\textbf{r}^{\ensuremath{'}})}{\partial r_a \partial r^{\ensuremath{'}}_b} \phi^{*}_{i}(\textbf{r})\phi^{*}_{j}(\textbf{r}^{\ensuremath{'}})d\textbf{r} d\textbf{r}^{\ensuremath{'}} }}\,,
\end{equation}
and
\begin{equation} \label{Eq. 6}
    M^{ij,E}_{ab} = \int{\int{\phi_{i}(\textbf{r})\phi_{j}(\textbf{r}^{\ensuremath{'}})  \frac{\partial^{2}G(\textbf{r},\textbf{r}^{\ensuremath{'}})}{\partial r_a \partial r^{\ensuremath{'}}_b} \phi^{*}_{i}(\textbf{r}^{\ensuremath{'}})\phi^{*}_{j}(\textbf{r}) d\textbf{r} d\textbf{r}^{\ensuremath{'}}}}\,.
\end{equation}
Equation~\ref{Eq. 5} and Eq. \ref{Eq. 6} can be rewritten as 
\begin{equation} \label{Eq. 7}
    M^{ij,D}_{ab}= \int{\int{\frac{\partial (\phi_{i}(\textbf{r})\phi^{*}_{i}(\textbf{r}))}{\partial r_a} G(\textbf{r},\textbf{r}^{\ensuremath{'}}) \frac{\partial (\phi_{j}(\textbf{r}^{\ensuremath{'}})\phi^{*}_{j}(\textbf{r}^{\ensuremath{'}}))}{\partial r^{\ensuremath{'}}_b}d\textbf{r} d\textbf{r}^{\ensuremath{'}}}}\,,
\end{equation}
and
\begin{equation} \label{Eq. 8}
    M^{ij,E}_{ab} = \int{\int{\frac{\partial( \phi_{i}(\textbf{r})\phi^{*}_{j}(\textbf{r}))}{\partial r_a} G(\textbf{r},\textbf{r}^{\ensuremath{'}}) \frac{\partial (\phi^{*}_{i}(\textbf{r}^{\ensuremath{'}})\phi_{j}(\textbf{r}^{\ensuremath{'}}))}{\partial r^{\ensuremath{'}}_b}d\textbf{r} d\textbf{r}^{\ensuremath{'}}}}\,.
\end{equation}
While the equivalence of Eqs. 5-6 with Eqs. 7-8 is trivial to see for molecular systems (using integration by parts), showing the equivalence for periodic systems requires a more complex manipulation (see Supplemental Material). 

In order to evaluate the double integrals in Eq. \ref{Eq. 7} and Eq. \ref{Eq. 8}, we note that the kernel of extended interactions is the Green's function of $-\frac{1}{4\pi}\nabla^{2}$, and we take recourse to the solution of the Poisson equation. Thus, we obtain,
\begin{equation} \label{Eq. 9}
    M^{ij,D}_{ab} = \int{\frac{\partial (\phi_{i}(\textbf{r})\phi^{*}_{i}(\textbf{r}))}{\partial r_a}  \Lambda^{jj,D}_{b} (\textbf{r}) d\textbf{r}}
\end{equation}
and
\begin{equation} \label{Eq. 10}
    M^{ij,E}_{ab} = \int{\frac{\partial (\phi_{i}(\textbf{r})\phi^{*}_{j}(\textbf{r}))}{\partial r_a} \Lambda^{ij,E}_{b}(\textbf{r}) d\textbf{r}}\,,
\end{equation}
where $\nabla^{2}\Lambda^{jj,D}_{b}(\textbf{r})=-4\pi\frac{\partial ( \phi_{j}(\textbf{r})\phi^{*}_{j}(\textbf{r}))}{\partial r_b}$ and $\nabla^{2}\Lambda^{ij,E}_{b}(\textbf{r})=-4\pi\frac{\partial (\phi^{*}_{i}(\textbf{r})\phi_{j}(\textbf{r}))}{\partial r_b}$. Thus, finally, the $D$-tensor can be expressed as 
\begin{equation} \label{Eq. 11}
    D_{ab} = \frac{1}{2S(2S-1)} \frac{\mu_0}{4\pi} (\gamma_e \hbar)^2 \sum_{i < j}^{occ.} \chi_{ij} ( M^{ij,D}_{ab} -M^{ij,E}_{ab})\,.
\end{equation}

The computationally expensive part of the D-tensor calculation involves the solution of Poisson problems, which are solved on the same FE mesh that represents the KS wavefunctions. However, this computation is embarrassingly parallel over the pairs of orbitals $\phi_{i}$ and $\phi_{j}$. We note that, unlike the $A$-tensor, the dipole-dipole integral entering the $D$-tensor expression (Eq. \ref{Eq. 4}) does not explicitly depend on the nuclear coordinates, and thus we expect the $D$-tensor to be less sensitive to the cusps in the spin density at the nuclei. Therefore, a \textit{p}-refinement is ideal to systematically improve the accuracy in the calculation of $D$.

\subsection{Electric Field Gradient Tensor}
The nuclear quadrupole interaction $P$-tensor is directly related to the electric field gradient (EFG) V-tensor. We denote the nuclear quadrupole moment by $Q$ and the quantum number ($a$ component) of the nuclear spin as $I$ ($I_a$); the nuclear quadrupole Hamiltonian is given by~\cite{Schweiger2001}
\begin{equation}\label{Eq. 12}
\begin{split}
    H_Q &= \bm{I} \cdot \bm{P} \cdot \bm{I} \\
        &= \frac{eQ}{6I(2I-1)} \sum_{a, b} V_{ab} \left[ \frac{3}{2}(I_a I_b + I_b I_a) - \delta_{ab} I(I+1) \right],
\end{split}
\end{equation}
where the EFG $V$-tensor is the second derivative of the electrostatic potential at the nucleus:
\begin{equation}\label{Eq. 13}
\begin{split}
    V_{ab} &= [ \nabla_a \nabla_b V(\bm{r}) ] |_{\bm{r} = \bm{r}_N} \\
    &= \left\{ \nabla_a \nabla_b \left[ - \int d\bm{r}' \frac{ n(\bm{r}') }{ |\bm{r} - \bm{r}'| } + \sum_{I \neq N} \frac{ Z_I }{ |\bm{r} - \bm{r}_I| } \right] \right\} \Biggr|_{\bm{r} = \bm{r}_N}
\end{split}
\end{equation}
Here $n$ is the electron density (defined as positive), and $Z_I$ and $\bm{r}_I$ are the charge and position of the $I^\text{th}$ nucleus in the system, respectively.

Calculation of the nuclear contribution to the $V$-tensor (second term in Eq. \ref{Eq. 13}) is trivial, and only requires the knowledge of the nuclear charges and the respective positions of the nuclei. We note that the electronic contribution to the $V$-tensor is given by the Hessian of the electrostatic potential. To this end, from a converged self-consistent DFT calculation, we extract the Hartree potential and compute the Hessian at the FE quadrature points. By construction, every nucleus is on an FE node in the FE mesh. Thus, the value of the Hessian at each nucleus is obtained via a projection of the quadrature point values to nodal value. As the $V$-tensor involves point-wise second-order derivatives, a careful convergence study of both \textit{h} and \textit{p} refinement is required.

\section{Results and discussions} \label{Sec 3}
We carried out calculations of spin Hamiltonian parameters for a series of molecules/radicals and the nitrogen-vacancy (NV) center in diamond. For the calculation of the NV center, the -1 charge state was considered, which is the most relevant charge state for NV-based quantum information processing. A 64-atom supercell of diamond and $\Gamma$-point sampling of the Brillouin zone were used. In the following discussion of $A$ and $V$-tensors of the NV center, we focus on the nitrogen atom and the three carbon atoms with dangling bonds (DB). All calculations were performed with the PBE functional~\cite{Perdew1996}. When treating charged systems we included a neutralizing jellium background. All structures were optimized with plane-wave DFT using the \texttt{QUANTUM ESPRESSO} code~\cite{Giannozzi2009} and the same structures were used for all-electron calculations.

All-electron FE calculations were performed with the \texttt{DFT-FE} code~\cite{dftfe} using adaptive real-space meshes. The tensor elements were converged with respect to the FE basis through $h$ and $p$ refinements,  within 1-2 MHz for the $A$-tensor, $5 \times 10^{-4} \, \text{cm}^{-1}$ for the $D$-tensor and 0.05 a.u. for the $V$-tensor of molecules. Convergence of the spin Hamiltonian parameters for the NV center with respect to the FE basis is presented later in the discussion. 

In order to verify our FE results, we also performed PW-based calculations for all systems and GTO-based calculations for molecules. PW calculations of the $A$ and $V$ tensors were carried out with the \texttt{GIPAW} code using the GIPAW pseudopotentials (PP)~\cite{Ceresoli}. PW calculations of the $D$-tensor were conducted with two different PP: GIPAW and ONCV~\cite{Schlipf2015}. We followed the numerical method in Ref. ~\onlinecite{Rayson2008} to evaluate Eq. \ref{Eq. 4} in reciprocal space, using normalized pseudo-wavefunctions~\cite{Ivady2014, Falk2014, Seo2017, Whiteley2018} (without PAW reconstructions) from the \texttt{QUANTUM ESPRESSO} code. A kinetic energy cutoff of 200 Ry was used for PW calculations of molecules; for the NV center we used 100 Ry for computational efficiency. GTO calculations of $A$, $D$ and $P$ tensors were carried out with the \texttt{ORCA} code~\cite{Neese2012}. Two Gaussian basis sets were considered: EPR-III~\cite{Rega1996} and IGLO-III~\cite{Kutzelnigg1990}, both of which are designed for an accurate representation of core electrons. We also tested a series of general-purpose basis sets from Dunning and co-workers (cc-pVDZ, cc-pVTZ, cc-pVQZ and cc-pV5Z)~\cite{Dunning1989}, but we found that the values of $A$, $D$ and $P$ tensors converge poorly as a function of the basis set and the poor convergence prevented any meaningful extrapolation to the complete basis set limit. We present GTO results obtained with cc- basis sets in the Supplemental Material.

Table-I and Table-II show the isotropic (Fermi contact) and spin dipolar component of the $A$-tensor for several molecules (CN, BO, AlO, NH) and the NV center. Due to the symmetry of the systems considered here, the dipolar $A$-tensor has only one independent component (except for DB carbons in the NV center). Denoting the principal values of the dipolar $A$-tensor as $A^{\text{dip}}_{11}, A^{\text{dip}}_{22}, A^{\text{dip}}_{33}$ ($|A^{\text{dip}}_{11}| = |A^{\text{dip}}_{22}| = \frac{1}{2} |A^{\text{dip}}_{33}|$), we show $A^{\text{dip}}_{33}$ in Table-II. In PW calculations we tested three different treatments of core relaxation  (Slater exchange-only, exchange-only and exchange-correlation) implemented in the \texttt{GIPAW} code~\cite{Bahramy2007}. Experimental values are also shown in the Tables for reference. We note that all of the results presented here, in addition to numerical errors which are quantified and discussed in detail below, suffer from systematic errors introduced by the use of a specific, approximate exchange-correlation functional, the PBE functional. A previous study has shown that more advanced functionals, such as certain hybrid and meta-GGA functionals, may improve the agreement with experiments, relative to GGA functionals, for the $A$-tensor of small radicals and transition metal complexes~\cite{Kossmann2007}. However, there is yet no consensus on which functional is the most accurate one, in general, for the calculation of the $A$-tensor or other spin Hamiltonian parameters. 

\begin{table*}[ht]
\caption{Isotropic hyperfine tensor (see Eq. \ref{Eq. 2}) (MHz) computed by DFT with finite-element (FE), Gaussian-type orbital (GTO) and plane-wave (PW) basis sets. For PW calculations three different treatments of the core-relaxation effect are considered, which include Slater exchange (Slater-X), exchange (X), and exchange+correlation (XC) in the perturbative potential for the calculation of spin densities at the core region.}
\begin{tabular}{ccccccccc}
\hline
  System &              Atom &       FE & GTO(EPR-III) & GTO(IGLO-III) & PW(Slater-X) &   PW(X) &   PW(XC) &            Exp \\
\hline
  $\text{C}\text{N}$ ($S = \frac{1}{2}$) &   $^{13}\text{C}$ &   504.21 &       500.50 &        509.63 &       539.55 &  536.04 &   566.57 &                      588 [\onlinecite{Easley1970}] \\
                                         &   $^{14}\text{N}$ &   -12.81 &       -12.47 &        -12.25 &       -14.87 &  -15.15 &   -12.43 &                      -13 [\onlinecite{Easley1970}] \\
\hline
  $\text{B}\text{O}$ ($S = \frac{1}{2}$) &   $^{11}\text{B}$ &  1007.71 &       998.17 &       1002.34 &       983.68 &  980.02 &  1009.31 &                   1027 [\onlinecite{Tanimoto1986}] \\
                                         &   $^{17}\text{O}$ &    -7.34 &        -7.18 &         -7.26 &        -7.83 &   -8.13 &    -7.24 &                                              \\
\hline
 $\text{Al}\text{O}$ ($S = \frac{1}{2}$) &  $^{27}\text{Al}$ &   590.80 &              &        646.95 &       564.18 &  560.09 &   626.93 &                        766 [\onlinecite{Knight1971}] \\
                                         &   $^{17}\text{O}$ &    11.47 &              &         12.20 &       -15.22 &  -14.78 &   -22.99 &                                              \\
\hline
            $\text{N}\text{H}$ ($S = 1$) &   $^{14}\text{N}$ &    11.27 &        10.20 &          9.77 &        24.24 &   22.00 &    33.60 &                        20 [\onlinecite{Weltner1983}] \\
                                         &    $^{1}\text{H}$ &   -53.52 &       -53.13 &        -47.74 &       -51.60 &  -51.60 &   -51.60 &                       -70 [\onlinecite{Weltner1983}] \\
\hline
                    Diamond NV ($S = 1$) &   $^{14}\text{N}$ &    -2.32 &              &               &        -2.60 &   -2.60 &    -2.56 &  2.23 [\onlinecite{He1993}], -2.51 [\onlinecite{Felton2009}], -2.53 [\onlinecite{Yavkin2016}] \\
                                         &   DB $^{13}\text{C}$ &    98.72 &              &               &       100.27 &   99.05 &   108.51 &                                       146.7 [\onlinecite{Felton2009}] \\
\hline
\end{tabular}
\label{hyperfine_iso_table}
\end{table*}

\begin{table*}[ht]
\caption{Spin dipolar hyperfine tensor (see Eq. \ref{Eq. 3}) (MHz) computed by DFT with finite-element (FE), Gaussian-type orbital (GTO) and plane-wave (PW) basis sets. The eigenvalue with the largest absolute value is shown.}
\begin{tabular}{ccccccc}
\hline
System &              Atom &       FE & GTO(EPR-III) & GTO(IGLO-III) &       PW &           Exp \\
\hline
  $\text{C}\text{N}$ ($S = \frac{1}{2}$) &   $^{13}\text{C}$ &   115.33 &       118.47 &        117.43 &   124.20 &                      89.9 [\onlinecite{Easley1970}] \\
                                         &   $^{14}\text{N}$ &    44.51 &        42.62 &         42.40 &    45.25 &                      30.8 [\onlinecite{Easley1970}] \\
\hline
  $\text{B}\text{O}$ ($S = \frac{1}{2}$) &   $^{11}\text{B}$ &    53.71 &        53.38 &         53.76 &    55.37 &                  54.254 [\onlinecite{Tanimoto1986}] \\
                                         &   $^{17}\text{O}$ &   -47.83 &       -46.47 &        -45.97 &   -51.55 &                                             \\
\hline
 $\text{Al}\text{O}$ ($S = \frac{1}{2}$) &  $^{27}\text{Al}$ &   114.23 &              &        111.67 &   112.51 &                       106 [\onlinecite{Knight1971}] \\
                                         &   $^{17}\text{O}$ &  -122.65 &              &       -116.42 &  -127.57 &                                             \\
\hline
            $\text{N}\text{H}$ ($S = 1$) &   $^{14}\text{N}$ &   -47.87 &       -45.82 &        -46.01 &   -49.59 &                      -46 [\onlinecite{Weltner1983}] \\
                                         &    $^{1}\text{H}$ &    58.08 &        58.50 &         59.92 &    58.02 &                       60 [\onlinecite{Weltner1983}] \\
\hline
                    Diamond NV ($S = 1$) &   $^{14}\text{N}$ &    -0.07 &              &               &    -0.05 &  -0.13 [\onlinecite{He1993}], 0.37 [\onlinecite{Felton2009}],
                    0.33 [\onlinecite{Yavkin2016}] \\
                                         &   DB $^{13}\text{C}$ &    54.87 &              &               &    58.34 &                       52.9 [\onlinecite{Felton2009}] \\
\hline
\end{tabular}
\label{hyperfine_dip_table}
\end{table*}

We found that in general, GTO results obtained with EPR-III and IGLO-III basis sets are similar, with a mean absolute deviation (MAD) of 3.2 (0.6) MHz for $A^{\text{iso}}$ ($A^{\text{dip}}$) for the systems considered here. FE and GTO results agree well: the MAD between FE and GTO@EPR-III results is 2.5 (1.5) MHz for $A^{\text{iso}}$ ($A^{\text{dip}}$). However, for the Al atom in AlO, FE and GTO@IGLO-III yield different values of $A^{\text{iso}}$ by 56 MHz (9\%). We expect the difference to originate from inaccuracies of the IGLO-III basis set used in GTO calculation; for example, we found that GTO calculations using different cc- basis sets yield large variations, between 580 to 520 MHz, for the $A^{\text{iso}}$ value of Al (see Supplemental Materials). Overall, the agreement between FE and GTO results serves as a verification of our FE implementation for the calculation of the $A$-tensor. We note that EPR-III and IGLO-III sets are specialized GTO basis designed for spin Hamiltonian parameter calculations, and they are not available for all elements (for instance, an EPR-III basis set for Al is not available). FE-based calculations, on the other hand, can be performed for any element in the periodic table and the results can be systematically converged with respect to the basis set. 

We found that PW calculations agree well with all-electron FE and GTO calculations for $A^{\text{dip}}$, while they deviate slightly for $A^{\text{iso}}$. For $A^{\text{dip}}$, the MAD between FE and PW results is 2.7 MHz, while the MAD for $A^{\text{iso}}$ ranges from 13-17 MHz depending on the treatment of core relaxation in PW calculations. Notably, in the case of the AlO molecule, PW calculations predicted a different sign for the $A^{\text{iso}}$ of the O atom compared to all-electron FE and GTO calculations. 

PW and FE calculations for the NV center yielded qualitatively similar values for $A^{\text{iso}}$ and $A^{\text{dip}}$ for both nitrogen and DB carbons. The larger value of $A^{\text{iso}}$ compared to $A^{\text{dip}}$ for the nitrogen atom reveals a strong \textit{s} character of the spin density on the nitrogen. The spin density on the DB carbons has instead a significant \textit{p}-type contribution as revealed by the comparable values of $A^{\text{iso}}$ and $A^{\text{dip}}$. There is a sizable difference between DFT results and experimental values for the $A^{\text{iso}}$ of DB carbons (30\%), which might be due to the use of a small (64-atom) supercell for the NV center.

In Table-III we present the computed zero field splitting $D$-tensor for several spin-triplet molecules/radicals (O\textsubscript{2}, CH\textsubscript{2}, NH, C\textsubscript{5}H\textsubscript{5}\textsuperscript{+}) as well as for the NV center. We report the scalar parameter $D = \frac{3}{2} D_{33}$, where $D_{11}, D_{22}, D_{33}$ are principal values of the $D$-tensor such that $|D_{11}| \leq |D_{22}| \leq |D_{33}|$. For low symmetry systems such as the CH\textsubscript{2} carbene, we additionally report the scalar parameter $E = \frac{1}{2} (D_{11} - D_{22})$.

\begin{table*}[ht]
\caption{The spin-spin component of the zero-field splitting tensor (see Eq. \ref{Eq. 4}) ($\text{cm}^{-1}$) computed by DFT with finite-element (FE), Gaussian-type orbital (GTO) and plane-wave (PW) basis sets. Scalar parameters $D = \frac{3}{2} D_{33}$ are reported. Scalar parameters $E = \frac{1}{2} (D_{22} - D_{11})$, if non-zero, are reported in brackets. }
\begin{tabular}{ccccccc}
\hline
                          System &             FE &   GTO(EPR-III) &  GTO(IGLO-III) &      PW(GIPAW) &       PW(ONCV) &            Exp \\
\hline
                 $\text{O}_{2}$ ($S = 1$) &          1.894 &          1.893 &          1.843 &          1.642 &          1.695 &              3.960 [\onlinecite{Huber1979}]$^{a}$ \\
         $\text{C}\text{H}_{2}$ ($S = 1$) &  0.894 (0.051) &  0.895 (0.052) &  0.895 (0.052) &  0.896 (0.052) &  0.908 (0.049) &  0.760 (0.068) [\onlinecite{Wasserman1971}] \\
             $\text{N}\text{H}$ ($S = 1$) &          1.861 &          1.862 &          1.857 &          1.795 &          1.815 &              1.860 [\onlinecite{Dixon1959}] \\
 $\text{C}_{5}\text{H}_{5}^{+}$ ($S = 1$) &          0.123 &          0.120 &          0.110 &          0.123 &          0.123 &           0.187 [\onlinecite{Saunders1973}] \\
                     Diamond NV ($S = 1$) &          0.100 &                &                &          0.101 &          0.100 &            0.096 [\onlinecite{Loubser1978}] \\
\hline
\end{tabular}

$^{a}$ Experimental value for O\textsubscript{2} is dominated by the spin-orbit component. The spin-spin component is estimated to be $1.57 \, \text{cm}^{-1}$ by \textit{ab initio} wavefunction-based calculations~\cite{Sinnecker2006}.
\label{zfs_table}
\end{table*}

Overall, GTO results show a weak dependence on the basis set, with a MAD of 0.016 $\text{cm}^{-1}$ between values obtained with EPR-III and IGLO-III basis sets. PW calculations show a weak dependence on the chosen pseudopotential, with a MAD of 0.017 $\text{cm}^{-1}$ between ONCV and GIPAW results. Similar to the case of the $A$-tensor, GTO and FE results agree well, with a MAD of 0.001 $\text{cm}^{-1}$ between FE and GTO@EPR-III values. Due to the use of pseudo-wavefunctions for the evaluation of Eq. \ref{Eq. 4} and the lack of PAW reconstruction, PW results deviate from all-electron ones, with a MAD of 0.064 $\text{cm}^{-1}$ between FE and PW@GIPAW values. For the case of the NV center, results from FE, PW and experiments appear to be in good agreement.

Table-IV summarizes the electric field gradient $V$-tensor for several closed-shell molecules (HCN, NCCN, N\textsubscript{2}, H\textsubscript{2}O) and for the NV center. Following the convention of the NQR spectroscopy literature, we report the quadrupole coupling constants $eQV_{33}$, where $V_{11}, V_{22}, V_{33}$ are principal values of the $V$-tensor such that $|V_{11}| \leq |V_{22}| \leq |V_{33}|$. When computing the $eQV_{33}$ values in Table-IV we considered isotopes with non-zero nuclear quadrupole moment $Q$, and the values of $Q$ are obtained from Ref. \onlinecite{WebElements}. For low symmetry systems, we additionally report $\eta = \left| \frac{V_{22} - V_{11}}{V_{33}} \right|$. 

\begin{table*}[ht]
\caption{Quadrupole coupling constants $eQ V_{33}$ (see Eq. \ref{Eq. 13}) (MHz) computed by DFT with finite-element (FE), Gaussian-type orbital (GTO) and plane-wave (PW) basis sets. The scalar parameters $\eta = \left| \frac{V_{22} - V_{11}}{V_{33}} \right|$, if non-zero, are reported in brackets.}
\begin{tabular}{ccccccc}
\hline
                            System &             Atom &            FE &  GTO(EPR-III) & GTO(IGLO-III) &            PW &            Exp \\
\hline
         $\text{H}\text{C}\text{N}$ ($S = 0$) &   $^{2}\text{H}$ &          0.20 &          0.21 &          0.22 &          0.20 &                                    \\
                                              &  $^{11}\text{C}$ &         -2.95 &         -2.87 &         -2.77 &         -3.73 &                                    \\
                                              &  $^{14}\text{N}$ &         -4.72 &         -4.85 &         -4.79 &         -5.07 &        -4.02 [\onlinecite{Latosinska2003}] \\
\hline
 $\text{N}\text{C}\text{C}\text{N}$ ($S = 0$) &  $^{14}\text{N}$ &         -4.72 &         -4.80 &         -4.75 &         -4.99 &        -4.27 [\onlinecite{Latosinska2003}] \\
                                              &  $^{11}\text{C}$ &         -1.84 &         -1.95 &         -1.76 &         -2.72 &                                    \\
\hline
                     $\text{N}_{2}$ ($S = 0$) &  $^{14}\text{N}$ &         -5.08 &         -5.52 &         -5.46 &         -5.77 &        -4.65 [\onlinecite{Latosinska2003}] \\
\hline
             $\text{H}_{2}\text{O}$ ($S = 0$) &   $^{2}\text{H}$ &   0.31 (0.12) &   0.31 (0.13) &   0.32 (0.13) &   0.30 (0.13) &   0.31 (0.14) [\onlinecite{Verhoeven1969}] \\
                                              &  $^{17}\text{O}$ &  10.10 (0.81) &  10.69 (0.74) &  10.90 (0.74) &  10.85 (0.72) &  10.17 (0.75) [\onlinecite{Verhoeven1969}] \\
\hline
                         Diamond NV ($S = 1$) &  $^{14}\text{N}$ &         -7.30 &               &               &         -7.47 &            -6.68 [\onlinecite{Felton2009}]$^{a}$ \\
                                              &  DB $^{11}\text{C}$ &   1.96 (0.03) &               &               &  -1.77 (0.10) &                                    \\
\hline
\end{tabular}

$^{a}$ Computed from $P_{\parallel}$ parameter reported in Ref. \onlinecite{Felton2009} by $eQV_{33} = 4I(2I - 1) P_{\parallel} / 3$.
\label{efg_table}
\end{table*}

Unlike the $A$ tensor, which depends on charge density differences, the $V$-tensor depends on the absolute value of the charge density and thus it is more sensitive to the details of the electronic structure. Differences are indeed observed for GTO calculations with different basis sets (MAD = 0.09 MHz), as well as between GTO and FE calculations (MAD = 0.18 MHz between FE and GTO@EPR-III). PW results significantly deviate from all-electron GTO and FE results, with a MAD of 0.76 MHz between FE and PW values. In the case of the NV center, PW and FE yield similar nuclear quadrupole coupling for nitrogen, in qualitative agreement with experiment, while for DB carbons the predicted $V_{33}$ values using PW and FE have opposite signs.

Finally, to demonstrate the convergence of the FE results with respect to the basis set, in Table-V we show A-, D- and V-tensors for the NV center computed with different FE polynomial degrees. We denote calculations with $n^{\text{th}}$-order polynomials as FE$n$. For the A-tensor and V-tensor calculations a mesh size of 0.1 Bohr was used surrounding the nuclei, while for the D-tensor calculation the mesh size was 0.5 Bohr. We see in Table-V that our results for the A-tensor are well converged at the FE6 level, as indicated by the small difference (less than $3\%$) between FE5 and FE6 results. Similarly, D-tensor values are well converged at the FE5 level. The numerical value of the $V$ tensor is sensitive to the details of the electronic wavefunctions around the nuclei, as mentioned previously, and its  convergence is indeed more challenging compared to that of the A- and D-tensors. We performed V-tensor calculations with polynomial degrees up to 7. At the FE7 level, most of the computed V tensor elements are converged within $10\%$, based on  asymptotic estimates obtained by power law extrapolations.

\begin{table*}[ht]
\caption{The principle values of $A$ (top, in MHz), $D$ (center, in $\text{cm}^{-1}$), and $V$ (bottom, in a.u.) tensors for the NV center computed by FE-based DFT using different finite-element polynomial degrees.}
\begin{tabular}{@{}c|cccc|cccc@{}}
\toprule
    & \multicolumn{4}{c|}{$^{14}$N}                                                                                                                     & \multicolumn{4}{c}{DB $^{13}$C}                                                                                                                     \\ \hline
    & A\SPSB{dip}{11} & A\SPSB{dip}{22} & A\SPSB{dip}{33} & A\textsuperscript{iso} & A\SPSB{dip}{11} & A\SPSB{dip}{22} & A\SPSB{dip}{33} & A\textsuperscript{iso} \\ \hline
FE3 & 0.033                              & 0.033                              & -0.066                             & -2.362                     & -27.303                            & -27.592                            & 54.896                             & 100.492                    \\
FE4 & 0.034                              & 0.034                              & -0.067                             & -2.307                     & -27.199                            & -27.654                            & 54.853                             & 99.708                     \\
FE5 & 0.035                              & 0.035                              & -0.070                             & -2.319                     & -27.189                            & -27.664                            & 54.854                             & 99.016                     \\
FE6 & 0.035                              & 0.035                              & -0.070                             & -2.316                     & -27.171                            & -27.696                            & 54.867                             & 98.721                     \\ \hline
\end{tabular}
\\[4pt]
\begin{tabular}{@{}c|ccc@{}}
\toprule
    & D\textsubscript{11} & D\textsubscript{22} & D\textsubscript{33} \\ \hline
FE3 & -0.0327    & -0.0327    & 0.0654     \\
FE4 & -0.0321    & -0.0321    & 0.0642     \\
FE5 & -0.0329    & -0.0329    & 0.0658     \\ \hline
\end{tabular}
\\[4pt]
\begin{tabular}{@{}c|ccc|ccc@{}}
\toprule
    & \multicolumn{3}{c|}{N}             & \multicolumn{3}{c}{DB C}             \\ \hline
    & V\textsubscript{11} & V\textsubscript{22} & V\textsubscript{33} & V\textsubscript{11} & V\textsubscript{22} & V\textsubscript{33} \\ \hline
FE5 &   0.865   & 0.865    & -1.731     & -0.033     & -0.127    &  0.160    \\
FE6 &   0.804   & 0.804    & -1.609     & -0.081     & -0.136    &  0.217   \\
FE7 &   0.761   & 0.761    & -1.520     & -0.122     & -0.129    &  0.251   \\ \hline
\end{tabular}
\end{table*}

\section{Conclusions}
In this work, we presented an approach to compute spin Hamiltonian parameters based on DFT, which uses all-electron calculations and finite element basis sets to solve the Kohn-Sham equations. The approach can be applied to both solids and molecules and offers the important advantage of straightforward convergence of the calculations with respect to the basis set, which can be systematically achieved by refinement of the finite element basis. 

We reported calculations of the Fermi contact and dipolar component of the $A$-tensor, the spin-spin component of the $D$-tensor and the nuclear quadrupole $P$-tensor for several molecules and for the NV center in diamond. We presented detailed comparisons of results obtained using FE, GTO and PW basis. For molecules, we showed that all-electron results obtained with FE basis sets are in good agreement with those obtained with GTO basis sets.

The approach introduced in our work represents the first step towards building a robust protocol for the first-principles prediction of various spin Hamiltonian parameters based on finite element density functional theory. There are multiple prospects of future work in this direction, both in terms of the level of physics and computational efficiency. It is important to extend the current formalism to include relativistic effects since proper treatment of scalar relativistic effects will be crucial for accurate calculations of spin Hamiltonian parameters of heavy elements. The ability to include spin-orbit coupling effects will also allow for the computation of additional spin Hamiltonian parameters, including the $g$-tensor and the spin-orbit component of the $A$ and $D$-tensor. Further, it would be interesting to develop and test more advanced density functionals, such as meta-GGAs and hybrid functionals, and to establish which functional performs better, compared to experiments. With regards to the computational efficiency, the FE basis functions can be enriched using compactly supported precomputed enrichment functions~\cite{Kanungo2017}, which will drastically reduce the computational cost, while providing systematic convergence. Finally, we plan to utilize a combination of all-electron and pseudopotential based calculation under the same framework, where certain atoms of interest are treated at all-electron level and other atoms are treated using pseudopotential approximation, which will enable the computation of spin Hamiltonian parameters in systems involving thousands of atoms. \\ [5pt]

We thank P. Motamarri for useful discussions on various aspects of the numerical implementation. This work was supported by MICCoM as part of the Computational Materials Sciences Program funded by the U.S. Department of Energy, Office of Science, Basic Energy Sciences, Materials Sciences and Engineering Division through Argonne National Laboratory, under contract number DE-AC02-06CH11357. V.G. also gratefully acknowledges the support of the Department of Energy, Office of Basic Energy Sciences, through grant number DE-SC0017380, under the auspices of which the computational framework for FE-based all-electron DFT calculations was developed. K.G. acknowledges the computational resources from the University of Michigan through the Flux computing platform. H. M. acknowledges computational resources from the University of Chicago Research Computing Center. \\ [5pt]

K.G. and H.M. have contributed equally to this work.

\newpage
\bibliography{main}

\begin{thebibliography}{62}%
\makeatletter
\providecommand \@ifxundefined [1]{%
 \@ifx{#1\undefined}
}%
\providecommand \@ifnum [1]{%
 \ifnum #1\expandafter \@firstoftwo
 \else \expandafter \@secondoftwo
 \fi
}%
\providecommand \@ifx [1]{%
 \ifx #1\expandafter \@firstoftwo
 \else \expandafter \@secondoftwo
 \fi
}%
\providecommand \natexlab [1]{#1}%
\providecommand \enquote  [1]{``#1''}%
\providecommand \bibnamefont  [1]{#1}%
\providecommand \bibfnamefont [1]{#1}%
\providecommand \citenamefont [1]{#1}%
\providecommand \href@noop [0]{\@secondoftwo}%
\providecommand \href [0]{\begingroup \@sanitize@url \@href}%
\providecommand \@href[1]{\@@startlink{#1}\@@href}%
\providecommand \@@href[1]{\endgroup#1\@@endlink}%
\providecommand \@sanitize@url [0]{\catcode `\\12\catcode `\$12\catcode
  `\&12\catcode `\#12\catcode `\^12\catcode `\_12\catcode `\%12\relax}%
\providecommand \@@startlink[1]{}%
\providecommand \@@endlink[0]{}%
\providecommand \url  [0]{\begingroup\@sanitize@url \@url }%
\providecommand \@url [1]{\endgroup\@href {#1}{\urlprefix }}%
\providecommand \urlprefix  [0]{URL }%
\providecommand \Eprint [0]{\href }%
\providecommand \doibase [0]{http://dx.doi.org/}%
\providecommand \selectlanguage [0]{\@gobble}%
\providecommand \bibinfo  [0]{\@secondoftwo}%
\providecommand \bibfield  [0]{\@secondoftwo}%
\providecommand \translation [1]{[#1]}%
\providecommand \BibitemOpen [0]{}%
\providecommand \bibitemStop [0]{}%
\providecommand \bibitemNoStop [0]{.\EOS\space}%
\providecommand \EOS [0]{\spacefactor3000\relax}%
\providecommand \BibitemShut  [1]{\csname bibitem#1\endcsname}%
\let\auto@bib@innerbib\@empty
\bibitem [{\citenamefont {\ifmmode \check{Z}\else
  \v{Z}\fi{}uti\ifmmode~\acute{c}\else \'{c}\fi{}}\ \emph
  {et~al.}(2004)\citenamefont {\ifmmode \check{Z}\else
  \v{Z}\fi{}uti\ifmmode~\acute{c}\else \'{c}\fi{}}, \citenamefont {Fabian},\
  and\ \citenamefont {Das~Sarma}}]{Zutic2004}%
  \BibitemOpen
  \bibfield  {author} {\bibinfo {author} {\bibfnamefont {I.}~\bibnamefont
  {\ifmmode \check{Z}\else \v{Z}\fi{}uti\ifmmode~\acute{c}\else \'{c}\fi{}}},
  \bibinfo {author} {\bibfnamefont {J.}~\bibnamefont {Fabian}}, \ and\ \bibinfo
  {author} {\bibfnamefont {S.}~\bibnamefont {Das~Sarma}},\ }\href@noop {}
  {\bibfield  {journal} {\bibinfo  {journal} {Rev. Mod. Phys.}\ }\textbf
  {\bibinfo {volume} {76}},\ \bibinfo {pages} {323} (\bibinfo {year}
  {2004})}\BibitemShut {NoStop}%
\bibitem [{\citenamefont {Weber}\ \emph {et~al.}(2010)\citenamefont {Weber},
  \citenamefont {Koehl}, \citenamefont {Varley}, \citenamefont {Janotti},
  \citenamefont {Buckley}, \citenamefont {Van~de Walle},\ and\ \citenamefont
  {Awschalom}}]{Weber2010}%
  \BibitemOpen
  \bibfield  {author} {\bibinfo {author} {\bibfnamefont {J.}~\bibnamefont
  {Weber}}, \bibinfo {author} {\bibfnamefont {W.}~\bibnamefont {Koehl}},
  \bibinfo {author} {\bibfnamefont {J.}~\bibnamefont {Varley}}, \bibinfo
  {author} {\bibfnamefont {A.}~\bibnamefont {Janotti}}, \bibinfo {author}
  {\bibfnamefont {B.}~\bibnamefont {Buckley}}, \bibinfo {author} {\bibfnamefont
  {C.}~\bibnamefont {Van~de Walle}}, \ and\ \bibinfo {author} {\bibfnamefont
  {D.~D.}\ \bibnamefont {Awschalom}},\ }\href@noop {} {\bibfield  {journal}
  {\bibinfo  {journal} {Proc. Natl. Acad. Sci. U. S. A.}\ }\textbf {\bibinfo
  {volume} {107}},\ \bibinfo {pages} {8513} (\bibinfo {year}
  {2010})}\BibitemShut {NoStop}%
\bibitem [{\citenamefont {Guo}\ \emph {et~al.}(2018)\citenamefont {Guo},
  \citenamefont {Day}, \citenamefont {Chen}, \citenamefont {Tong},
  \citenamefont {Mansikkam{\"a}ki},\ and\ \citenamefont {Layfield}}]{Guo2018}%
  \BibitemOpen
  \bibfield  {author} {\bibinfo {author} {\bibfnamefont {F.-S.}\ \bibnamefont
  {Guo}}, \bibinfo {author} {\bibfnamefont {B.~M.}\ \bibnamefont {Day}},
  \bibinfo {author} {\bibfnamefont {Y.-C.}\ \bibnamefont {Chen}}, \bibinfo
  {author} {\bibfnamefont {M.-L.}\ \bibnamefont {Tong}}, \bibinfo {author}
  {\bibfnamefont {A.}~\bibnamefont {Mansikkam{\"a}ki}}, \ and\ \bibinfo
  {author} {\bibfnamefont {R.~A.}\ \bibnamefont {Layfield}},\ }\href@noop {} {\
  \textbf {\bibinfo {volume} {362}},\ \bibinfo {pages} {1400} (\bibinfo {year}
  {2018})}\BibitemShut {NoStop}%
\bibitem [{\citenamefont {Koehl}\ \emph {et~al.}(2015)\citenamefont {Koehl},
  \citenamefont {Seo}, \citenamefont {Galli},\ and\ \citenamefont
  {Awschalom}}]{Koehl2015}%
  \BibitemOpen
  \bibfield  {author} {\bibinfo {author} {\bibfnamefont {W.~F.}\ \bibnamefont
  {Koehl}}, \bibinfo {author} {\bibfnamefont {H.}~\bibnamefont {Seo}}, \bibinfo
  {author} {\bibfnamefont {G.}~\bibnamefont {Galli}}, \ and\ \bibinfo {author}
  {\bibfnamefont {D.~D.}\ \bibnamefont {Awschalom}},\ }\href@noop {} {\bibfield
   {journal} {\bibinfo  {journal} {MRS Bull.}\ }\textbf {\bibinfo {volume}
  {40}},\ \bibinfo {pages} {1146} (\bibinfo {year} {2015})}\BibitemShut
  {NoStop}%
\bibitem [{\citenamefont {Schweiger}\ and\ \citenamefont
  {Jeschke}(2001)}]{Schweiger2001}%
  \BibitemOpen
  \bibfield  {author} {\bibinfo {author} {\bibfnamefont {A.}~\bibnamefont
  {Schweiger}}\ and\ \bibinfo {author} {\bibfnamefont {G.}~\bibnamefont
  {Jeschke}},\ }\href@noop {} {\emph {\bibinfo {title} {Principles of pulse
  electron paramagnetic resonance}}}\ (\bibinfo  {publisher} {Oxford University
  Press},\ \bibinfo {year} {2001})\BibitemShut {NoStop}%
\bibitem [{\citenamefont {Harriman}(2013)}]{Harriman2013}%
  \BibitemOpen
  \bibfield  {author} {\bibinfo {author} {\bibfnamefont {J.~E.}\ \bibnamefont
  {Harriman}},\ }\href@noop {} {\emph {\bibinfo {title} {Theoretical
  Foundations of Electron Spin Resonance}}}\ (\bibinfo  {publisher} {Academic
  press},\ \bibinfo {year} {2013})\BibitemShut {NoStop}%
\bibitem [{\citenamefont {Abragam}\ and\ \citenamefont
  {Bleaney}(2013)}]{Abragam2013}%
  \BibitemOpen
  \bibfield  {author} {\bibinfo {author} {\bibfnamefont {A.}~\bibnamefont
  {Abragam}}\ and\ \bibinfo {author} {\bibfnamefont {B.}~\bibnamefont
  {Bleaney}},\ }\href@noop {} {\emph {\bibinfo {title} {Electron Paramagnetic
  Resonance of Transition Ions}}}\ (\bibinfo  {publisher} {Oxford University
  Press},\ \bibinfo {year} {2013})\BibitemShut {NoStop}%
\bibitem [{\citenamefont {Weil}\ and\ \citenamefont {Bolton}(2007)}]{Weil2007}%
  \BibitemOpen
  \bibfield  {author} {\bibinfo {author} {\bibfnamefont {J.~A.}\ \bibnamefont
  {Weil}}\ and\ \bibinfo {author} {\bibfnamefont {J.~R.}\ \bibnamefont
  {Bolton}},\ }\href@noop {} {\emph {\bibinfo {title} {Electron paramagnetic
  resonance: elementary theory and practical applications}}}\ (\bibinfo
  {publisher} {John Wiley \& Sons},\ \bibinfo {year} {2007})\BibitemShut
  {NoStop}%
\bibitem [{\citenamefont {Falk}\ \emph {et~al.}(2014)\citenamefont {Falk},
  \citenamefont {Klimov}, \citenamefont {Buckley}, \citenamefont {Iv{\'a}dy},
  \citenamefont {Abrikosov}, \citenamefont {Calusine}, \citenamefont {Koehl},
  \citenamefont {Gali},\ and\ \citenamefont {Awschalom}}]{Falk2014}%
  \BibitemOpen
  \bibfield  {author} {\bibinfo {author} {\bibfnamefont {A.~L.}\ \bibnamefont
  {Falk}}, \bibinfo {author} {\bibfnamefont {P.~V.}\ \bibnamefont {Klimov}},
  \bibinfo {author} {\bibfnamefont {B.~B.}\ \bibnamefont {Buckley}}, \bibinfo
  {author} {\bibfnamefont {V.}~\bibnamefont {Iv{\'a}dy}}, \bibinfo {author}
  {\bibfnamefont {I.~A.}\ \bibnamefont {Abrikosov}}, \bibinfo {author}
  {\bibfnamefont {G.}~\bibnamefont {Calusine}}, \bibinfo {author}
  {\bibfnamefont {W.~F.}\ \bibnamefont {Koehl}}, \bibinfo {author}
  {\bibfnamefont {{\'A}.}~\bibnamefont {Gali}}, \ and\ \bibinfo {author}
  {\bibfnamefont {D.~D.}\ \bibnamefont {Awschalom}},\ }\href@noop {} {\bibfield
   {journal} {\bibinfo  {journal} {Phys. Rev. Lett.}\ }\textbf {\bibinfo
  {volume} {112}},\ \bibinfo {pages} {187601} (\bibinfo {year}
  {2014})}\BibitemShut {NoStop}%
\bibitem [{\citenamefont {Whiteley}\ \emph {et~al.}()\citenamefont {Whiteley},
  \citenamefont {Wolfowicz}, \citenamefont {Anderson}, \citenamefont
  {Bourassa}, \citenamefont {Ma}, \citenamefont {Ye}, \citenamefont {Koolstra},
  \citenamefont {Satzinger}, \citenamefont {Holt}, \citenamefont {Heremans},
  \citenamefont {Cleland}, \citenamefont {Schuster}, \citenamefont {Galli},\
  and\ \citenamefont {Awschalom}}]{Whiteley2018}%
  \BibitemOpen
  \bibfield  {author} {\bibinfo {author} {\bibfnamefont {S.~J.}\ \bibnamefont
  {Whiteley}}, \bibinfo {author} {\bibfnamefont {G.}~\bibnamefont {Wolfowicz}},
  \bibinfo {author} {\bibfnamefont {C.~P.}\ \bibnamefont {Anderson}}, \bibinfo
  {author} {\bibfnamefont {A.}~\bibnamefont {Bourassa}}, \bibinfo {author}
  {\bibfnamefont {H.}~\bibnamefont {Ma}}, \bibinfo {author} {\bibfnamefont
  {M.}~\bibnamefont {Ye}}, \bibinfo {author} {\bibfnamefont {G.}~\bibnamefont
  {Koolstra}}, \bibinfo {author} {\bibfnamefont {K.~J.}\ \bibnamefont
  {Satzinger}}, \bibinfo {author} {\bibfnamefont {M.~V.}\ \bibnamefont {Holt}},
  \bibinfo {author} {\bibfnamefont {F.~J.}\ \bibnamefont {Heremans}}, \bibinfo
  {author} {\bibfnamefont {A.~N.}\ \bibnamefont {Cleland}}, \bibinfo {author}
  {\bibfnamefont {D.~I.}\ \bibnamefont {Schuster}}, \bibinfo {author}
  {\bibfnamefont {G.}~\bibnamefont {Galli}}, \ and\ \bibinfo {author}
  {\bibfnamefont {D.~D.}\ \bibnamefont {Awschalom}},\ }\href@noop {} {\bibinfo
  {journal} {arXiv:1804.10996}\ }\BibitemShut {NoStop}%
\bibitem [{\citenamefont {Van~de Walle}\ and\ \citenamefont
  {Bl\"ochl}(1993)}]{Vandewalle1993}%
  \BibitemOpen
\bibfield  {journal} {  }\bibfield  {author} {\bibinfo {author} {\bibfnamefont
  {C.~G.}\ \bibnamefont {Van~de Walle}}\ and\ \bibinfo {author} {\bibfnamefont
  {P.~E.}\ \bibnamefont {Bl\"ochl}},\ }\href@noop {} {\bibfield  {journal}
  {\bibinfo  {journal} {Phys. Rev. B}\ }\textbf {\bibinfo {volume} {47}},\
  \bibinfo {pages} {4244} (\bibinfo {year} {1993})}\BibitemShut {NoStop}%
\bibitem [{\citenamefont {Bl\"ugel}\ \emph {et~al.}(1987)\citenamefont
  {Bl\"ugel}, \citenamefont {Akai}, \citenamefont {Zeller},\ and\ \citenamefont
  {Dederichs}}]{Blugel1987}%
  \BibitemOpen
  \bibfield  {author} {\bibinfo {author} {\bibfnamefont {S.}~\bibnamefont
  {Bl\"ugel}}, \bibinfo {author} {\bibfnamefont {H.}~\bibnamefont {Akai}},
  \bibinfo {author} {\bibfnamefont {R.}~\bibnamefont {Zeller}}, \ and\ \bibinfo
  {author} {\bibfnamefont {P.~H.}\ \bibnamefont {Dederichs}},\ }\href@noop {}
  {\bibfield  {journal} {\bibinfo  {journal} {Phys. Rev. B}\ }\textbf {\bibinfo
  {volume} {35}},\ \bibinfo {pages} {3271} (\bibinfo {year}
  {1987})}\BibitemShut {NoStop}%
\bibitem [{\citenamefont {Pickard}\ and\ \citenamefont
  {Mauri}(2001)}]{Pickard2001}%
  \BibitemOpen
  \bibfield  {author} {\bibinfo {author} {\bibfnamefont {C.~J.}\ \bibnamefont
  {Pickard}}\ and\ \bibinfo {author} {\bibfnamefont {F.}~\bibnamefont
  {Mauri}},\ }\href@noop {} {\bibfield  {journal} {\bibinfo  {journal} {Phys.
  Rev. B}\ }\textbf {\bibinfo {volume} {63}},\ \bibinfo {pages} {245101}
  (\bibinfo {year} {2001})}\BibitemShut {NoStop}%
\bibitem [{\citenamefont {Pickard}\ and\ \citenamefont
  {Mauri}(2002)}]{Pickard2002}%
  \BibitemOpen
  \bibfield  {author} {\bibinfo {author} {\bibfnamefont {C.~J.}\ \bibnamefont
  {Pickard}}\ and\ \bibinfo {author} {\bibfnamefont {F.}~\bibnamefont
  {Mauri}},\ }\href@noop {} {\bibfield  {journal} {\bibinfo  {journal} {Phys.
  Rev. Lett.}\ }\textbf {\bibinfo {volume} {88}},\ \bibinfo {pages} {086403}
  (\bibinfo {year} {2002})}\BibitemShut {NoStop}%
\bibitem [{\citenamefont {Bahramy}\ \emph {et~al.}(2007)\citenamefont
  {Bahramy}, \citenamefont {Sluiter},\ and\ \citenamefont
  {Kawazoe}}]{Bahramy2007}%
  \BibitemOpen
  \bibfield  {author} {\bibinfo {author} {\bibfnamefont {M.~S.}\ \bibnamefont
  {Bahramy}}, \bibinfo {author} {\bibfnamefont {M.~H.}\ \bibnamefont
  {Sluiter}}, \ and\ \bibinfo {author} {\bibfnamefont {Y.}~\bibnamefont
  {Kawazoe}},\ }\href@noop {} {\bibfield  {journal} {\bibinfo  {journal} {Phys.
  Rev. B}\ }\textbf {\bibinfo {volume} {76}},\ \bibinfo {pages} {035124}
  (\bibinfo {year} {2007})}\BibitemShut {NoStop}%
\bibitem [{\citenamefont {Rayson}\ and\ \citenamefont
  {Briddon}(2008)}]{Rayson2008}%
  \BibitemOpen
  \bibfield  {author} {\bibinfo {author} {\bibfnamefont {M.}~\bibnamefont
  {Rayson}}\ and\ \bibinfo {author} {\bibfnamefont {P.}~\bibnamefont
  {Briddon}},\ }\href@noop {} {\bibfield  {journal} {\bibinfo  {journal} {Phys.
  Rev. B}\ }\textbf {\bibinfo {volume} {77}},\ \bibinfo {pages} {035119}
  (\bibinfo {year} {2008})}\BibitemShut {NoStop}%
\bibitem [{\citenamefont {Bodrog}\ and\ \citenamefont
  {Gali}(2013)}]{Bodrog2013}%
  \BibitemOpen
  \bibfield  {author} {\bibinfo {author} {\bibfnamefont {Z.}~\bibnamefont
  {Bodrog}}\ and\ \bibinfo {author} {\bibfnamefont {A.}~\bibnamefont {Gali}},\
  }\href@noop {} {\bibfield  {journal} {\bibinfo  {journal} {J. Phys.: Condens.
  Matter}\ }\textbf {\bibinfo {volume} {26}},\ \bibinfo {pages} {015305}
  (\bibinfo {year} {2013})}\BibitemShut {NoStop}%
\bibitem [{\citenamefont {Biktagirov}\ \emph {et~al.}(2018)\citenamefont
  {Biktagirov}, \citenamefont {Schmidt},\ and\ \citenamefont
  {Gerstmann}}]{Biktagirov2018}%
  \BibitemOpen
  \bibfield  {author} {\bibinfo {author} {\bibfnamefont {T.}~\bibnamefont
  {Biktagirov}}, \bibinfo {author} {\bibfnamefont {W.~G.}\ \bibnamefont
  {Schmidt}}, \ and\ \bibinfo {author} {\bibfnamefont {U.}~\bibnamefont
  {Gerstmann}},\ }\href@noop {} {\bibfield  {journal} {\bibinfo  {journal}
  {Phys. Rev. B}\ }\textbf {\bibinfo {volume} {97}},\ \bibinfo {pages} {115135}
  (\bibinfo {year} {2018})}\BibitemShut {NoStop}%
\bibitem [{\citenamefont {Olsen}\ \emph {et~al.}(2002)\citenamefont {Olsen},
  \citenamefont {Christiansen}, \citenamefont {Hemmingsen}, \citenamefont
  {Sauer},\ and\ \citenamefont {Mikkelsen}}]{Olsen2002}%
  \BibitemOpen
  \bibfield  {author} {\bibinfo {author} {\bibfnamefont {L.}~\bibnamefont
  {Olsen}}, \bibinfo {author} {\bibfnamefont {O.}~\bibnamefont {Christiansen}},
  \bibinfo {author} {\bibfnamefont {L.}~\bibnamefont {Hemmingsen}}, \bibinfo
  {author} {\bibfnamefont {S.~P.}\ \bibnamefont {Sauer}}, \ and\ \bibinfo
  {author} {\bibfnamefont {K.~V.}\ \bibnamefont {Mikkelsen}},\ }\href@noop {}
  {\bibfield  {journal} {\bibinfo  {journal} {J. Chem. Phys.}\ }\textbf
  {\bibinfo {volume} {116}},\ \bibinfo {pages} {1424} (\bibinfo {year}
  {2002})}\BibitemShut {NoStop}%
\bibitem [{\citenamefont {Latosi{\'n}ska}(2003)}]{Latosinska2003}%
  \BibitemOpen
  \bibfield  {author} {\bibinfo {author} {\bibfnamefont {J.}~\bibnamefont
  {Latosi{\'n}ska}},\ }\href@noop {} {\bibfield  {journal} {\bibinfo  {journal}
  {Int. J. Quantum Chem.}\ }\textbf {\bibinfo {volume} {91}},\ \bibinfo {pages}
  {284} (\bibinfo {year} {2003})}\BibitemShut {NoStop}%
\bibitem [{\citenamefont {Sinnecker}\ and\ \citenamefont
  {Neese}(2006)}]{Sinnecker2006}%
  \BibitemOpen
  \bibfield  {author} {\bibinfo {author} {\bibfnamefont {S.}~\bibnamefont
  {Sinnecker}}\ and\ \bibinfo {author} {\bibfnamefont {F.}~\bibnamefont
  {Neese}},\ }\href@noop {} {\bibfield  {journal} {\bibinfo  {journal} {J.
  Phys. Chem. A}\ }\textbf {\bibinfo {volume} {110}},\ \bibinfo {pages} {12267}
  (\bibinfo {year} {2006})}\BibitemShut {NoStop}%
\bibitem [{\citenamefont {Neese}(2005)}]{Neese2005}%
  \BibitemOpen
  \bibfield  {author} {\bibinfo {author} {\bibfnamefont {F.}~\bibnamefont
  {Neese}},\ }\href@noop {} {\bibfield  {journal} {\bibinfo  {journal} {J.
  Chem. Phys.}\ }\textbf {\bibinfo {volume} {122}},\ \bibinfo {pages} {034107}
  (\bibinfo {year} {2005})}\BibitemShut {NoStop}%
\bibitem [{\citenamefont {Reviakine}\ \emph {et~al.}(2006)\citenamefont
  {Reviakine}, \citenamefont {Arbuznikov}, \citenamefont {Tremblay},
  \citenamefont {Remenyi}, \citenamefont {Malkina}, \citenamefont {Malkin},\
  and\ \citenamefont {Kaupp}}]{Reviakine2006}%
  \BibitemOpen
  \bibfield  {author} {\bibinfo {author} {\bibfnamefont {R.}~\bibnamefont
  {Reviakine}}, \bibinfo {author} {\bibfnamefont {A.~V.}\ \bibnamefont
  {Arbuznikov}}, \bibinfo {author} {\bibfnamefont {J.-C.}\ \bibnamefont
  {Tremblay}}, \bibinfo {author} {\bibfnamefont {C.}~\bibnamefont {Remenyi}},
  \bibinfo {author} {\bibfnamefont {O.~L.}\ \bibnamefont {Malkina}}, \bibinfo
  {author} {\bibfnamefont {V.~G.}\ \bibnamefont {Malkin}}, \ and\ \bibinfo
  {author} {\bibfnamefont {M.}~\bibnamefont {Kaupp}},\ }\href@noop {}
  {\bibfield  {journal} {\bibinfo  {journal} {J. Chem. Phys.}\ }\textbf
  {\bibinfo {volume} {125}},\ \bibinfo {pages} {054110} (\bibinfo {year}
  {2006})}\BibitemShut {NoStop}%
\bibitem [{\citenamefont {Kossmann}\ \emph {et~al.}(2007)\citenamefont
  {Kossmann}, \citenamefont {Kirchner},\ and\ \citenamefont
  {Neese}}]{Kossmann2007}%
  \BibitemOpen
  \bibfield  {author} {\bibinfo {author} {\bibfnamefont {S.}~\bibnamefont
  {Kossmann}}, \bibinfo {author} {\bibfnamefont {B.}~\bibnamefont {Kirchner}},
  \ and\ \bibinfo {author} {\bibfnamefont {F.}~\bibnamefont {Neese}},\
  }\href@noop {} {\bibfield  {journal} {\bibinfo  {journal} {Mol. Phys.}\
  }\textbf {\bibinfo {volume} {105}},\ \bibinfo {pages} {2049} (\bibinfo {year}
  {2007})}\BibitemShut {NoStop}%
\bibitem [{\citenamefont {Neese}(2007)}]{Neese2007}%
  \BibitemOpen
  \bibfield  {author} {\bibinfo {author} {\bibfnamefont {F.}~\bibnamefont
  {Neese}},\ }\href@noop {} {\bibfield  {journal} {\bibinfo  {journal} {J.
  Chem. Phys.}\ }\textbf {\bibinfo {volume} {127}},\ \bibinfo {pages} {164112}
  (\bibinfo {year} {2007})}\BibitemShut {NoStop}%
\bibitem [{\citenamefont {Sayfutyarova}\ and\ \citenamefont
  {Chan}(2018)}]{Sayfutyarova2018}%
  \BibitemOpen
  \bibfield  {author} {\bibinfo {author} {\bibfnamefont {E.~R.}\ \bibnamefont
  {Sayfutyarova}}\ and\ \bibinfo {author} {\bibfnamefont {G.~K.-L.}\
  \bibnamefont {Chan}},\ }\href@noop {} {\bibfield  {journal} {\bibinfo
  {journal} {J. Chem. Phys.}\ }\textbf {\bibinfo {volume} {148}},\ \bibinfo
  {pages} {184103} (\bibinfo {year} {2018})}\BibitemShut {NoStop}%
\bibitem [{\citenamefont {Sugisaki}\ \emph {et~al.}(2009)\citenamefont
  {Sugisaki}, \citenamefont {Toyota}, \citenamefont {Sato}, \citenamefont
  {Shiomi}, \citenamefont {Kitagawa},\ and\ \citenamefont
  {Takui}}]{Sugisaki2009}%
  \BibitemOpen
  \bibfield  {author} {\bibinfo {author} {\bibfnamefont {K.}~\bibnamefont
  {Sugisaki}}, \bibinfo {author} {\bibfnamefont {K.}~\bibnamefont {Toyota}},
  \bibinfo {author} {\bibfnamefont {K.}~\bibnamefont {Sato}}, \bibinfo {author}
  {\bibfnamefont {D.}~\bibnamefont {Shiomi}}, \bibinfo {author} {\bibfnamefont
  {M.}~\bibnamefont {Kitagawa}}, \ and\ \bibinfo {author} {\bibfnamefont
  {T.}~\bibnamefont {Takui}},\ }\href@noop {} {\bibfield  {journal} {\bibinfo
  {journal} {Chem. Phys. Lett.}\ }\textbf {\bibinfo {volume} {477}},\ \bibinfo
  {pages} {369} (\bibinfo {year} {2009})}\BibitemShut {NoStop}%
\bibitem [{\citenamefont {Bl\"ochl}(1994)}]{Blochl1994}%
  \BibitemOpen
  \bibfield  {author} {\bibinfo {author} {\bibfnamefont {P.~E.}\ \bibnamefont
  {Bl\"ochl}},\ }\href@noop {} {\bibfield  {journal} {\bibinfo  {journal}
  {Phys. Rev. B}\ }\textbf {\bibinfo {volume} {50}},\ \bibinfo {pages} {17953}
  (\bibinfo {year} {1994})}\BibitemShut {NoStop}%
\bibitem [{\citenamefont {Kadantsev}\ and\ \citenamefont
  {Ziegler}(2008)}]{Kadantsev2008}%
  \BibitemOpen
  \bibfield  {author} {\bibinfo {author} {\bibfnamefont {E.~S.}\ \bibnamefont
  {Kadantsev}}\ and\ \bibinfo {author} {\bibfnamefont {T.}~\bibnamefont
  {Ziegler}},\ }\href@noop {} {\bibfield  {journal} {\bibinfo  {journal} {J.
  Phys. Chem. A}\ }\textbf {\bibinfo {volume} {112}},\ \bibinfo {pages} {4521}
  (\bibinfo {year} {2008})}\BibitemShut {NoStop}%
\bibitem [{\citenamefont {Schwarz}\ and\ \citenamefont
  {Blaha}(2003)}]{Schwarz2003}%
  \BibitemOpen
  \bibfield  {author} {\bibinfo {author} {\bibfnamefont {K.}~\bibnamefont
  {Schwarz}}\ and\ \bibinfo {author} {\bibfnamefont {P.}~\bibnamefont
  {Blaha}},\ }\href@noop {} {\bibfield  {journal} {\bibinfo  {journal} {Comput.
  Mater. Sci.}\ }\textbf {\bibinfo {volume} {28}},\ \bibinfo {pages} {259 }
  (\bibinfo {year} {2003})}\BibitemShut {NoStop}%
\bibitem [{\citenamefont {Daalderop}\ \emph {et~al.}(1996)\citenamefont
  {Daalderop}, \citenamefont {Kelly},\ and\ \citenamefont
  {Schuurmans}}]{Daalderop1996}%
  \BibitemOpen
  \bibfield  {author} {\bibinfo {author} {\bibfnamefont {G.~H.~O.}\
  \bibnamefont {Daalderop}}, \bibinfo {author} {\bibfnamefont {P.~J.}\
  \bibnamefont {Kelly}}, \ and\ \bibinfo {author} {\bibfnamefont {M.~F.~H.}\
  \bibnamefont {Schuurmans}},\ }\href@noop {} {\bibfield  {journal} {\bibinfo
  {journal} {Phys. Rev. B}\ }\textbf {\bibinfo {volume} {53}},\ \bibinfo
  {pages} {14415} (\bibinfo {year} {1996})}\BibitemShut {NoStop}%
\bibitem [{\citenamefont {Overhof}\ and\ \citenamefont
  {Gerstmann}(2004)}]{Overhof2004}%
  \BibitemOpen
  \bibfield  {author} {\bibinfo {author} {\bibfnamefont {H.}~\bibnamefont
  {Overhof}}\ and\ \bibinfo {author} {\bibfnamefont {U.}~\bibnamefont
  {Gerstmann}},\ }\href@noop {} {\bibfield  {journal} {\bibinfo  {journal}
  {Phys. Rev. Lett.}\ }\textbf {\bibinfo {volume} {92}},\ \bibinfo {pages}
  {087602} (\bibinfo {year} {2004})}\BibitemShut {NoStop}%
\bibitem [{\citenamefont {Dovesi}\ \emph {et~al.}(2018)\citenamefont {Dovesi},
  \citenamefont {Erba}, \citenamefont {Orlando}, \citenamefont
  {Zicovich-Wilson}, \citenamefont {Civalleri}, \citenamefont {Maschio},
  \citenamefont {R{\'e}rat}, \citenamefont {Casassa}, \citenamefont {Baima},
  \citenamefont {Salustro},\ and\ \citenamefont {Kirtman}}]{Dovesi2018}%
  \BibitemOpen
  \bibfield  {author} {\bibinfo {author} {\bibfnamefont {R.}~\bibnamefont
  {Dovesi}}, \bibinfo {author} {\bibfnamefont {A.}~\bibnamefont {Erba}},
  \bibinfo {author} {\bibfnamefont {R.}~\bibnamefont {Orlando}}, \bibinfo
  {author} {\bibfnamefont {C.~M.}\ \bibnamefont {Zicovich-Wilson}}, \bibinfo
  {author} {\bibfnamefont {B.}~\bibnamefont {Civalleri}}, \bibinfo {author}
  {\bibfnamefont {L.}~\bibnamefont {Maschio}}, \bibinfo {author} {\bibfnamefont
  {M.}~\bibnamefont {R{\'e}rat}}, \bibinfo {author} {\bibfnamefont
  {S.}~\bibnamefont {Casassa}}, \bibinfo {author} {\bibfnamefont
  {J.}~\bibnamefont {Baima}}, \bibinfo {author} {\bibfnamefont
  {S.}~\bibnamefont {Salustro}}, \ and\ \bibinfo {author} {\bibfnamefont
  {B.}~\bibnamefont {Kirtman}},\ }\href@noop {} {\bibfield  {journal} {\bibinfo
   {journal} {Wiley Interdiscip. Rev.: Comput. Mol. Sci.}\ ,\ \bibinfo {pages}
  {e1360}} (\bibinfo {year} {2018})}\BibitemShut {NoStop}%
\bibitem [{\citenamefont {Motamarri}\ \emph {et~al.}(2013)\citenamefont
  {Motamarri}, \citenamefont {Nowak}, \citenamefont {Leiter}, \citenamefont
  {Knap},\ and\ \citenamefont {Gavini}}]{Motamarri2013}%
  \BibitemOpen
  \bibfield  {author} {\bibinfo {author} {\bibfnamefont {P.}~\bibnamefont
  {Motamarri}}, \bibinfo {author} {\bibfnamefont {M.~R.}\ \bibnamefont
  {Nowak}}, \bibinfo {author} {\bibfnamefont {K.}~\bibnamefont {Leiter}},
  \bibinfo {author} {\bibfnamefont {J.}~\bibnamefont {Knap}}, \ and\ \bibinfo
  {author} {\bibfnamefont {V.}~\bibnamefont {Gavini}},\ }\href@noop {}
  {\bibfield  {journal} {\bibinfo  {journal} {J. Comput. Phys.}\ }\textbf
  {\bibinfo {volume} {253}},\ \bibinfo {pages} {308} (\bibinfo {year}
  {2013})}\BibitemShut {NoStop}%
\bibitem [{dft()}]{dftfe}%
  \BibitemOpen
  \href@noop {} {\enquote {\bibinfo {title} {\texttt{DFT-FE} package},}\
  }\bibinfo {howpublished}
  {https://github.com/dftfeDevelopers/dftfe}\BibitemShut {NoStop}%
\bibitem [{\citenamefont {Brenner}\ and\ \citenamefont
  {Scott}(2009)}]{Brenner2008}%
  \BibitemOpen
  \bibfield  {author} {\bibinfo {author} {\bibfnamefont {S.~C.}\ \bibnamefont
  {Brenner}}\ and\ \bibinfo {author} {\bibfnamefont {R.}~\bibnamefont
  {Scott}},\ }\href@noop {} {\emph {\bibinfo {title} {The mathematical theory
  of finite element methods}}}\ (\bibinfo  {publisher} {Springer-Verlag, New
  York},\ \bibinfo {year} {2009})\BibitemShut {NoStop}%
\bibitem [{Web()}]{WebElements}%
  \BibitemOpen
  \href@noop {} {\enquote {\bibinfo {title} {Web of elements},}\ }\bibinfo
  {howpublished} {https://www.webelements.com/},\ \bibinfo {note} {accessed:
  Dec. 6, 2018}\BibitemShut {NoStop}%
\bibitem [{\citenamefont {McWeeny}(1961)}]{McWeeny1961}%
  \BibitemOpen
  \bibfield  {author} {\bibinfo {author} {\bibfnamefont {R.}~\bibnamefont
  {McWeeny}},\ }\href@noop {} {\bibfield  {journal} {\bibinfo  {journal} {Proc.
  R. Soc. London, Ser. A}\ }\textbf {\bibinfo {volume} {259}},\ \bibinfo
  {pages} {554} (\bibinfo {year} {1961})}\BibitemShut {NoStop}%
\bibitem [{\citenamefont {Perdew}\ \emph {et~al.}(1996)\citenamefont {Perdew},
  \citenamefont {Burke},\ and\ \citenamefont {Ernzerhof}}]{Perdew1996}%
  \BibitemOpen
  \bibfield  {author} {\bibinfo {author} {\bibfnamefont {J.~P.}\ \bibnamefont
  {Perdew}}, \bibinfo {author} {\bibfnamefont {K.}~\bibnamefont {Burke}}, \
  and\ \bibinfo {author} {\bibfnamefont {M.}~\bibnamefont {Ernzerhof}},\
  }\href@noop {} {\bibfield  {journal} {\bibinfo  {journal} {Phys. Rev. Lett.}\
  }\textbf {\bibinfo {volume} {77}},\ \bibinfo {pages} {3865} (\bibinfo {year}
  {1996})}\BibitemShut {NoStop}%
\bibitem [{\citenamefont {Giannozzi}\ \emph {et~al.}(2009)\citenamefont
  {Giannozzi}, \citenamefont {Baroni}, \citenamefont {Bonini}, \citenamefont
  {Calandra}, \citenamefont {Car}, \citenamefont {Cavazzoni}, \citenamefont
  {Ceresoli}, \citenamefont {Chiarotti}, \citenamefont {Cococcioni},
  \citenamefont {Dabo}, \citenamefont {Corso}, \citenamefont {de~Gironcoli},
  \citenamefont {Fabris}, \citenamefont {Fratesi}, \citenamefont {Gebauer},
  \citenamefont {Gerstmann}, \citenamefont {Gougoussis}, \citenamefont
  {Kokalj}, \citenamefont {Lazzeri}, \citenamefont {Martin-Samos},
  \citenamefont {Marzari}, \citenamefont {Mauri}, \citenamefont {Mazzarello},
  \citenamefont {Paolini}, \citenamefont {Pasquarello}, \citenamefont
  {Paulatto}, \citenamefont {Sbraccia}, \citenamefont {Scandolo}, \citenamefont
  {Sclauzero}, \citenamefont {Seitsonen}, \citenamefont {Smogunov},
  \citenamefont {Umari},\ and\ \citenamefont {Wentzcovitch}}]{Giannozzi2009}%
  \BibitemOpen
  \bibfield  {author} {\bibinfo {author} {\bibfnamefont {P.}~\bibnamefont
  {Giannozzi}}, \bibinfo {author} {\bibfnamefont {S.}~\bibnamefont {Baroni}},
  \bibinfo {author} {\bibfnamefont {N.}~\bibnamefont {Bonini}}, \bibinfo
  {author} {\bibfnamefont {M.}~\bibnamefont {Calandra}}, \bibinfo {author}
  {\bibfnamefont {R.}~\bibnamefont {Car}}, \bibinfo {author} {\bibfnamefont
  {C.}~\bibnamefont {Cavazzoni}}, \bibinfo {author} {\bibfnamefont
  {D.}~\bibnamefont {Ceresoli}}, \bibinfo {author} {\bibfnamefont {G.~L.}\
  \bibnamefont {Chiarotti}}, \bibinfo {author} {\bibfnamefont {M.}~\bibnamefont
  {Cococcioni}}, \bibinfo {author} {\bibfnamefont {I.}~\bibnamefont {Dabo}},
  \bibinfo {author} {\bibfnamefont {A.~D.}\ \bibnamefont {Corso}}, \bibinfo
  {author} {\bibfnamefont {S.}~\bibnamefont {de~Gironcoli}}, \bibinfo {author}
  {\bibfnamefont {S.}~\bibnamefont {Fabris}}, \bibinfo {author} {\bibfnamefont
  {G.}~\bibnamefont {Fratesi}}, \bibinfo {author} {\bibfnamefont
  {R.}~\bibnamefont {Gebauer}}, \bibinfo {author} {\bibfnamefont
  {U.}~\bibnamefont {Gerstmann}}, \bibinfo {author} {\bibfnamefont
  {C.}~\bibnamefont {Gougoussis}}, \bibinfo {author} {\bibfnamefont
  {A.}~\bibnamefont {Kokalj}}, \bibinfo {author} {\bibfnamefont
  {M.}~\bibnamefont {Lazzeri}}, \bibinfo {author} {\bibfnamefont
  {L.}~\bibnamefont {Martin-Samos}}, \bibinfo {author} {\bibfnamefont
  {N.}~\bibnamefont {Marzari}}, \bibinfo {author} {\bibfnamefont
  {F.}~\bibnamefont {Mauri}}, \bibinfo {author} {\bibfnamefont
  {R.}~\bibnamefont {Mazzarello}}, \bibinfo {author} {\bibfnamefont
  {S.}~\bibnamefont {Paolini}}, \bibinfo {author} {\bibfnamefont
  {A.}~\bibnamefont {Pasquarello}}, \bibinfo {author} {\bibfnamefont
  {L.}~\bibnamefont {Paulatto}}, \bibinfo {author} {\bibfnamefont
  {C.}~\bibnamefont {Sbraccia}}, \bibinfo {author} {\bibfnamefont
  {S.}~\bibnamefont {Scandolo}}, \bibinfo {author} {\bibfnamefont
  {G.}~\bibnamefont {Sclauzero}}, \bibinfo {author} {\bibfnamefont {A.~P.}\
  \bibnamefont {Seitsonen}}, \bibinfo {author} {\bibfnamefont {A.}~\bibnamefont
  {Smogunov}}, \bibinfo {author} {\bibfnamefont {P.}~\bibnamefont {Umari}}, \
  and\ \bibinfo {author} {\bibfnamefont {R.~M.}\ \bibnamefont {Wentzcovitch}},\
  }\href@noop {} {\bibfield  {journal} {\bibinfo  {journal} {J. Phys.: Condens.
  Matter}\ }\textbf {\bibinfo {volume} {21}},\ \bibinfo {pages} {395502}
  (\bibinfo {year} {2009})}\BibitemShut {NoStop}%
\bibitem [{Cer()}]{Ceresoli}%
  \BibitemOpen
  \href@noop {} {}\bibinfo {howpublished}
  {https://sites.google.com/site/dceresoli/pseudopotentials},\ \bibinfo {note}
  {accessed: Dec. 6, 2018}\BibitemShut {NoStop}%
\bibitem [{\citenamefont {Schlipf}\ and\ \citenamefont
  {Gygi}(2015)}]{Schlipf2015}%
  \BibitemOpen
  \bibfield  {author} {\bibinfo {author} {\bibfnamefont {M.}~\bibnamefont
  {Schlipf}}\ and\ \bibinfo {author} {\bibfnamefont {F.}~\bibnamefont {Gygi}},\
  }\href@noop {} {\bibfield  {journal} {\bibinfo  {journal} {Comput. Phys.
  Commun.}\ }\textbf {\bibinfo {volume} {196}},\ \bibinfo {pages} {36}
  (\bibinfo {year} {2015})}\BibitemShut {NoStop}%
\bibitem [{\citenamefont {Iv\'ady}\ \emph {et~al.}(2014)\citenamefont
  {Iv\'ady}, \citenamefont {Simon}, \citenamefont {Maze}, \citenamefont
  {Abrikosov},\ and\ \citenamefont {Gali}}]{Ivady2014}%
  \BibitemOpen
  \bibfield  {author} {\bibinfo {author} {\bibfnamefont {V.}~\bibnamefont
  {Iv\'ady}}, \bibinfo {author} {\bibfnamefont {T.}~\bibnamefont {Simon}},
  \bibinfo {author} {\bibfnamefont {J.~R.}\ \bibnamefont {Maze}}, \bibinfo
  {author} {\bibfnamefont {I.~A.}\ \bibnamefont {Abrikosov}}, \ and\ \bibinfo
  {author} {\bibfnamefont {A.}~\bibnamefont {Gali}},\ }\href@noop {} {\bibfield
   {journal} {\bibinfo  {journal} {Phys. Rev. B}\ }\textbf {\bibinfo {volume}
  {90}},\ \bibinfo {pages} {235205} (\bibinfo {year} {2014})}\BibitemShut
  {NoStop}%
\bibitem [{\citenamefont {Seo}\ \emph {et~al.}(2017)\citenamefont {Seo},
  \citenamefont {Ma}, \citenamefont {Govoni},\ and\ \citenamefont
  {Galli}}]{Seo2017}%
  \BibitemOpen
  \bibfield  {author} {\bibinfo {author} {\bibfnamefont {H.}~\bibnamefont
  {Seo}}, \bibinfo {author} {\bibfnamefont {H.}~\bibnamefont {Ma}}, \bibinfo
  {author} {\bibfnamefont {M.}~\bibnamefont {Govoni}}, \ and\ \bibinfo {author}
  {\bibfnamefont {G.}~\bibnamefont {Galli}},\ }\href@noop {} {\bibfield
  {journal} {\bibinfo  {journal} {Phys. Rev. Mater.}\ }\textbf {\bibinfo
  {volume} {1}},\ \bibinfo {pages} {075002} (\bibinfo {year}
  {2017})}\BibitemShut {NoStop}%
\bibitem [{\citenamefont {Neese}(2012)}]{Neese2012}%
  \BibitemOpen
  \bibfield  {author} {\bibinfo {author} {\bibfnamefont {F.}~\bibnamefont
  {Neese}},\ }\href@noop {} {\bibfield  {journal} {\bibinfo  {journal} {Wiley
  Interdiscip. Rev.: Comput. Mol. Sci.}\ }\textbf {\bibinfo {volume} {2}},\
  \bibinfo {pages} {73} (\bibinfo {year} {2012})}\BibitemShut {NoStop}%
\bibitem [{\citenamefont {Rega}\ \emph {et~al.}(1996)\citenamefont {Rega},
  \citenamefont {Cossi},\ and\ \citenamefont {Barone}}]{Rega1996}%
  \BibitemOpen
  \bibfield  {author} {\bibinfo {author} {\bibfnamefont {N.}~\bibnamefont
  {Rega}}, \bibinfo {author} {\bibfnamefont {M.}~\bibnamefont {Cossi}}, \ and\
  \bibinfo {author} {\bibfnamefont {V.}~\bibnamefont {Barone}},\ }\href@noop {}
  {\bibfield  {journal} {\bibinfo  {journal} {J. Chem. Phys.}\ }\textbf
  {\bibinfo {volume} {105}},\ \bibinfo {pages} {11060} (\bibinfo {year}
  {1996})}\BibitemShut {NoStop}%
\bibitem [{\citenamefont {Kutzelnigg}\ \emph {et~al.}(1990)\citenamefont
  {Kutzelnigg}, \citenamefont {Fleischer},\ and\ \citenamefont
  {Schindler}}]{Kutzelnigg1990}%
  \BibitemOpen
  \bibfield  {author} {\bibinfo {author} {\bibfnamefont {W.}~\bibnamefont
  {Kutzelnigg}}, \bibinfo {author} {\bibfnamefont {U.}~\bibnamefont
  {Fleischer}}, \ and\ \bibinfo {author} {\bibfnamefont {M.}~\bibnamefont
  {Schindler}},\ }in\ \href@noop {} {\emph {\bibinfo {booktitle} {Deuterium and
  shift calculation}}}\ (\bibinfo  {publisher} {Springer},\ \bibinfo {year}
  {1990})\ pp.\ \bibinfo {pages} {165--262}\BibitemShut {NoStop}%
\bibitem [{\citenamefont {Dunning~Jr}(1989)}]{Dunning1989}%
  \BibitemOpen
  \bibfield  {author} {\bibinfo {author} {\bibfnamefont {T.~H.}\ \bibnamefont
  {Dunning~Jr}},\ }\href@noop {} {\bibfield  {journal} {\bibinfo  {journal} {J.
  Chem. Phys.}\ }\textbf {\bibinfo {volume} {90}},\ \bibinfo {pages} {1007}
  (\bibinfo {year} {1989})}\BibitemShut {NoStop}%
\bibitem [{\citenamefont {Easley}\ and\ \citenamefont
  {Weltner~Jr}(1970)}]{Easley1970}%
  \BibitemOpen
  \bibfield  {author} {\bibinfo {author} {\bibfnamefont {W.~C.}\ \bibnamefont
  {Easley}}\ and\ \bibinfo {author} {\bibfnamefont {W.}~\bibnamefont
  {Weltner~Jr}},\ }\href@noop {} {\bibfield  {journal} {\bibinfo  {journal} {J.
  Chem. Phys.}\ }\textbf {\bibinfo {volume} {52}},\ \bibinfo {pages} {197}
  (\bibinfo {year} {1970})}\BibitemShut {NoStop}%
\bibitem [{\citenamefont {Tanimoto}\ \emph {et~al.}(1986)\citenamefont
  {Tanimoto}, \citenamefont {Saito},\ and\ \citenamefont
  {Hirota}}]{Tanimoto1986}%
  \BibitemOpen
  \bibfield  {author} {\bibinfo {author} {\bibfnamefont {M.}~\bibnamefont
  {Tanimoto}}, \bibinfo {author} {\bibfnamefont {S.}~\bibnamefont {Saito}}, \
  and\ \bibinfo {author} {\bibfnamefont {E.}~\bibnamefont {Hirota}},\
  }\href@noop {} {\bibfield  {journal} {\bibinfo  {journal} {J. Chem. Phys.}\
  }\textbf {\bibinfo {volume} {84}},\ \bibinfo {pages} {1210} (\bibinfo {year}
  {1986})}\BibitemShut {NoStop}%
\bibitem [{\citenamefont {Knight~Jr}\ and\ \citenamefont
  {Weltner~Jr}(1971)}]{Knight1971}%
  \BibitemOpen
  \bibfield  {author} {\bibinfo {author} {\bibfnamefont {L.}~\bibnamefont
  {Knight~Jr}}\ and\ \bibinfo {author} {\bibfnamefont {W.}~\bibnamefont
  {Weltner~Jr}},\ }\href@noop {} {\bibfield  {journal} {\bibinfo  {journal} {J.
  Chem. Phys.}\ }\textbf {\bibinfo {volume} {55}},\ \bibinfo {pages} {5066}
  (\bibinfo {year} {1971})}\BibitemShut {NoStop}%
\bibitem [{\citenamefont {Weltner}(1989)}]{Weltner1983}%
  \BibitemOpen
  \bibfield  {author} {\bibinfo {author} {\bibfnamefont {W.}~\bibnamefont
  {Weltner}},\ }\href@noop {} {\emph {\bibinfo {title} {Magnetic Atoms and
  Molecules}}}\ (\bibinfo  {publisher} {Dover Publications},\ \bibinfo {year}
  {1989})\BibitemShut {NoStop}%
\bibitem [{\citenamefont {He}\ \emph {et~al.}(1993)\citenamefont {He},
  \citenamefont {Manson},\ and\ \citenamefont {Fisk}}]{He1993}%
  \BibitemOpen
  \bibfield  {author} {\bibinfo {author} {\bibfnamefont {X.-F.}\ \bibnamefont
  {He}}, \bibinfo {author} {\bibfnamefont {N.~B.}\ \bibnamefont {Manson}}, \
  and\ \bibinfo {author} {\bibfnamefont {P.~T.}\ \bibnamefont {Fisk}},\
  }\href@noop {} {\bibfield  {journal} {\bibinfo  {journal} {Phys. Rev. B}\
  }\textbf {\bibinfo {volume} {47}},\ \bibinfo {pages} {8809} (\bibinfo {year}
  {1993})}\BibitemShut {NoStop}%
\bibitem [{\citenamefont {Felton}\ \emph {et~al.}(2009)\citenamefont {Felton},
  \citenamefont {Edmonds}, \citenamefont {Newton}, \citenamefont {Martineau},
  \citenamefont {Fisher}, \citenamefont {Twitchen},\ and\ \citenamefont
  {Baker}}]{Felton2009}%
  \BibitemOpen
  \bibfield  {author} {\bibinfo {author} {\bibfnamefont {S.}~\bibnamefont
  {Felton}}, \bibinfo {author} {\bibfnamefont {A.}~\bibnamefont {Edmonds}},
  \bibinfo {author} {\bibfnamefont {M.}~\bibnamefont {Newton}}, \bibinfo
  {author} {\bibfnamefont {P.}~\bibnamefont {Martineau}}, \bibinfo {author}
  {\bibfnamefont {D.}~\bibnamefont {Fisher}}, \bibinfo {author} {\bibfnamefont
  {D.}~\bibnamefont {Twitchen}}, \ and\ \bibinfo {author} {\bibfnamefont
  {J.}~\bibnamefont {Baker}},\ }\href@noop {} {\bibfield  {journal} {\bibinfo
  {journal} {Phys. Rev. B}\ }\textbf {\bibinfo {volume} {79}},\ \bibinfo
  {pages} {075203} (\bibinfo {year} {2009})}\BibitemShut {NoStop}%
\bibitem [{\citenamefont {Yavkin}\ \emph {et~al.}(2016)\citenamefont {Yavkin},
  \citenamefont {Mamin},\ and\ \citenamefont {Orlinskii}}]{Yavkin2016}%
  \BibitemOpen
  \bibfield  {author} {\bibinfo {author} {\bibfnamefont {B.}~\bibnamefont
  {Yavkin}}, \bibinfo {author} {\bibfnamefont {G.}~\bibnamefont {Mamin}}, \
  and\ \bibinfo {author} {\bibfnamefont {S.}~\bibnamefont {Orlinskii}},\
  }\href@noop {} {\bibfield  {journal} {\bibinfo  {journal} {J. Magn. Reson.}\
  }\textbf {\bibinfo {volume} {262}},\ \bibinfo {pages} {15 } (\bibinfo {year}
  {2016})}\BibitemShut {NoStop}%
\bibitem [{\citenamefont {Huber}(1979)}]{Huber1979}%
  \BibitemOpen
  \bibfield  {author} {\bibinfo {author} {\bibfnamefont {K.}~\bibnamefont
  {Huber}},\ }\href@noop {} {\emph {\bibinfo {title} {Molecular Spectra and
  Molecular Structure, Constants of Diatomic Molecules}}}\ (\bibinfo
  {publisher} {Springer},\ \bibinfo {year} {1979})\BibitemShut {NoStop}%
\bibitem [{\citenamefont {Wasserman}\ \emph {et~al.}(1971)\citenamefont
  {Wasserman}, \citenamefont {Hutton}, \citenamefont {Kuck},\ and\
  \citenamefont {Yager}}]{Wasserman1971}%
  \BibitemOpen
  \bibfield  {author} {\bibinfo {author} {\bibfnamefont {E.}~\bibnamefont
  {Wasserman}}, \bibinfo {author} {\bibfnamefont {R.}~\bibnamefont {Hutton}},
  \bibinfo {author} {\bibfnamefont {V.}~\bibnamefont {Kuck}}, \ and\ \bibinfo
  {author} {\bibfnamefont {W.}~\bibnamefont {Yager}},\ }\href@noop {}
  {\bibfield  {journal} {\bibinfo  {journal} {J. Chem. Phys.}\ }\textbf
  {\bibinfo {volume} {55}},\ \bibinfo {pages} {2593} (\bibinfo {year}
  {1971})}\BibitemShut {NoStop}%
\bibitem [{\citenamefont {Dixon}(1959)}]{Dixon1959}%
  \BibitemOpen
  \bibfield  {author} {\bibinfo {author} {\bibfnamefont {R.}~\bibnamefont
  {Dixon}},\ }\href@noop {} {\bibfield  {journal} {\bibinfo  {journal} {Can. J.
  Phys.}\ }\textbf {\bibinfo {volume} {37}},\ \bibinfo {pages} {1171} (\bibinfo
  {year} {1959})}\BibitemShut {NoStop}%
\bibitem [{\citenamefont {Saunders}\ \emph {et~al.}(1973)\citenamefont
  {Saunders}, \citenamefont {Berger}, \citenamefont {Jaffe}, \citenamefont
  {McBride}, \citenamefont {O'Neill}, \citenamefont {Breslow}, \citenamefont
  {Hoffmann}, \citenamefont {Perchonock}, \citenamefont {Wasserman},
  \citenamefont {Hutton},\ and\ \citenamefont {Kuck}}]{Saunders1973}%
  \BibitemOpen
  \bibfield  {author} {\bibinfo {author} {\bibfnamefont {M.}~\bibnamefont
  {Saunders}}, \bibinfo {author} {\bibfnamefont {R.}~\bibnamefont {Berger}},
  \bibinfo {author} {\bibfnamefont {A.}~\bibnamefont {Jaffe}}, \bibinfo
  {author} {\bibfnamefont {J.~M.}\ \bibnamefont {McBride}}, \bibinfo {author}
  {\bibfnamefont {J.}~\bibnamefont {O'Neill}}, \bibinfo {author} {\bibfnamefont
  {R.}~\bibnamefont {Breslow}}, \bibinfo {author} {\bibfnamefont {J.~M.}\
  \bibnamefont {Hoffmann}}, \bibinfo {author} {\bibfnamefont {C.}~\bibnamefont
  {Perchonock}}, \bibinfo {author} {\bibfnamefont {E.}~\bibnamefont
  {Wasserman}}, \bibinfo {author} {\bibfnamefont {R.~S.}\ \bibnamefont
  {Hutton}}, \ and\ \bibinfo {author} {\bibfnamefont {V.}~\bibnamefont
  {Kuck}},\ }\href@noop {} {\bibfield  {journal} {\bibinfo  {journal} {J. Am.
  Chem. Soc.}\ }\textbf {\bibinfo {volume} {95}},\ \bibinfo {pages} {3017}
  (\bibinfo {year} {1973})}\BibitemShut {NoStop}%
\bibitem [{\citenamefont {Loubser}\ and\ \citenamefont {van
  Wyk}(1978)}]{Loubser1978}%
  \BibitemOpen
  \bibfield  {author} {\bibinfo {author} {\bibfnamefont {J.}~\bibnamefont
  {Loubser}}\ and\ \bibinfo {author} {\bibfnamefont {J.}~\bibnamefont {van
  Wyk}},\ }\href@noop {} {\bibfield  {journal} {\bibinfo  {journal} {Rep. Prog.
  Phys.}\ }\textbf {\bibinfo {volume} {41}},\ \bibinfo {pages} {1201} (\bibinfo
  {year} {1978})}\BibitemShut {NoStop}%
\bibitem [{\citenamefont {Verhoeven}\ \emph {et~al.}(1969)\citenamefont
  {Verhoeven}, \citenamefont {Dymanus},\ and\ \citenamefont
  {Bluyssen}}]{Verhoeven1969}%
  \BibitemOpen
  \bibfield  {author} {\bibinfo {author} {\bibfnamefont {J.}~\bibnamefont
  {Verhoeven}}, \bibinfo {author} {\bibfnamefont {A.}~\bibnamefont {Dymanus}},
  \ and\ \bibinfo {author} {\bibfnamefont {H.}~\bibnamefont {Bluyssen}},\
  }\href@noop {} {\bibfield  {journal} {\bibinfo  {journal} {J. Chem. Phys.}\
  }\textbf {\bibinfo {volume} {50}},\ \bibinfo {pages} {3330} (\bibinfo {year}
  {1969})}\BibitemShut {NoStop}%
\bibitem [{\citenamefont {Kanungo}\ and\ \citenamefont
  {Gavini}(2017)}]{Kanungo2017}%
  \BibitemOpen
  \bibfield  {author} {\bibinfo {author} {\bibfnamefont {B.}~\bibnamefont
  {Kanungo}}\ and\ \bibinfo {author} {\bibfnamefont {V.}~\bibnamefont
  {Gavini}},\ }\href@noop {} {\bibfield  {journal} {\bibinfo  {journal} {Phys.
  Rev. B}\ }\textbf {\bibinfo {volume} {95}},\ \bibinfo {pages} {035112}
  (\bibinfo {year} {2017})}\BibitemShut {NoStop}%
\end{thebibliography}%

\end{document}